\documentclass[12pt]{article}

\usepackage{latexsym,amsbsy,amssymb,amsthm,amsmath,multirow}
\usepackage{natbib,graphicx,setspace,lscape,longtable}
\usepackage{array}
\usepackage{multirow}
\usepackage{threeparttable}
\usepackage[dvipsnames,usenames]{color}

\def\calS{\mathcal{S}} %The minimum size of screened covariate set to include all active predictors $\calS$
\def \p {{\rm Pr}}
\def \diag {\mathrm{diag}}
\def\.{{.}}
\newcommand{\trans}{^{T}}
\newcommand{\bSig}{\bf \Sigma}

\def\MDRSIS{MDR-SIS}

\def\MDRIS{MDR-ISIS}
\def\MDRSS{MDR-SSIS}

\newtheorem{theorem}{Theorem}[section]
\newtheorem{proposition}[theorem]{{\bf Proposition}}
\newtheorem{lemma}[theorem]{{\bf Lemma}}

\def\A{{\bf A}}
\def\C{{C}}
\def\T{{T}}
\def\D{{\bf D}}
\def\H{{\bf H}}
\def\B{{\bf B}}

\def\Y{Y}
\def\X{{\mathbf X}}
\def\e{{\mathbf e}}
\def\Z{{\mathbf Z}}
\def\M{{\bf M}}
\def\G{{\bf G}}
\def\W{{\mathbf W}}
\def\U{{\mathbf U}}
\def\V{{\mathbf V}}
\def\x{x}
\def\I{{\mathbf I}}
\def\0{{\textbf 0}}
\def\R{\mathbb{R}}

\def\cala{\mathcal{A}}
\def\calb{\mathcal{B}}
\def\cali{\mathcal{I}}
\def\cals{\mathcal{S}}
\def\calf{\mathcal{F}}

\def\ba{\mbox{\boldmath$\alpha$}}
\def\bfbeta{\mbox{\boldmath$\beta$}}
\def\bfLambda{\mbox{\boldmath$\Lambda$}}
\def\bfeta{\mbox{\boldmath$\eta$}}
\def\bfzeta{\mbox{\boldmath$\zeta$}}

\def\bfnu{\mbox{\boldmath$\nu$}}
\def\bfmu{\mbox{\boldmath$\mu$}}
\def\bfpsi{\mbox{\boldmath$\psi$}}
\def\sig{{\bf \Sigma}}

\newcommand{\indep}{\;\, \rule[0em]{.03em}{.67em} \hspace{-.25em}
\rule[0em]{.65em}{.03em} \hspace{-.25em}
\rule[0em]{.03em}{.67em}\;\,}

\newcommand{\spn}{\mathrm{Span}}
\newcommand{\var}{\mathrm{Var}}
\newcommand{\cov}{\mathrm{Cov}}

\def\eop{\hfill $\Box$ \\}
\numberwithin{equation}{section}
\makeatletter
\renewcommand\normalsize{%
	\@setfontsize\normalsize\@xiipt{17}%
	\abovedisplayskip 2\p@ \@plus2\p@ \@minus5\p@
	\abovedisplayshortskip \z@ \@plus3\p@
	\belowdisplayshortskip 10\p@ \@plus3\p@ \@minus3\p@
	\belowdisplayskip \abovedisplayskip
	\let\@listi\@listI}
\makeatother

\begin{document}
\def\spacingset#1{\renewcommand{\baselinestretch}%
	{#1}\small\normalsize} \spacingset{1}	

%\linenumbers
%
\title{\bf Sufficient variable screening  via directional regression with censored response}
\author{Menghao Xu$^1$, Zhou Yu$^1$, Jun Shao$^{1,2}$\\
	{\small $^{1}$ School of Statistics, East China Normal University, Shanghai, China} \\
	{\small 		$^{2}$Department of Statistics, University of Wisconsin-Madison, WI, 53706, USA}}
\maketitle

\begin{abstract}
	We in this paper propose a directional regression based approach for ultrahigh dimensional sufficient variable screening with censored responses. The new method is designed in a model-free manner and thus can be adapted to various complex model structures. Under some commonly used assumptions, we show that the proposed method enjoys the sure screening property when the dimension $p$ diverges at an exponential rate of the sample size $n$. To improve the marginal screening method, the corresponding iterative screening algorithm and stability screening algorithm are further equipped.
	We demonstrate the effectiveness of the proposed method through simulation studies and a real data analysis.
	
	\noindent{\it Key Words:} Sufficient dimension reduction, Sufficient Variable Selection; Sure independence screening; Ultrahigh dimensional covariates.
	%, sufficient dimension reduction.
\end{abstract}

\newpage
\spacingset{1.4}
%\csection{Introduction}
\section{Introduction}
\label{sec:intro}
Data sets collected in many contemporary scientific areas are ultrahigh dimensional and too complex to be analyzed through classical statistical methods.
%\begin{align}\label{model general}
%	\Y=f(\x_1,\ldots,\x_p,\varepsilon),
%\end{align}
%where $\varepsilon$ is an unobserved random error independent of $\X$ and $f$ is an unknown function on $\R^{p+1}$.
Consider data observed from a random sample of size $n$ from the 
distribution of $(\Y, \X)$, where 
$\Y$ is a scalar response,  $\X=(\x_1,\ldots,\x_p)\trans$
is a $p$-dimensional column vector of covariates, and 
the joint distribution of $( \Y, \X )$ is fully nonparametric. 
With an ultrahigh dimension $p >> n$, it is of great interest to identify 
$\cala \subset \{1,...,p\}$ such that $\X_\cala = \{ x_k: k \in \cala \}$ 
is truly related to the response. To fulfill the goal of model-free variable selection
based on the training data, 
\citet{Yin2015} introduced the concept of sufficient variable selection as finding the smallest covariate set $\X_{\cala}$ with $\cala \subset \{1,...,p\}$ satisfying 
\begin{align}\label{model svs}
\Y \indep \X \mid \X_{\cala},
\end{align}
where $\indep$ stands for independence and $\mid $ stands for conditioning. For convenience, in what follows we name both  $\cali \subseteq \{1,...,p\}$ and  $\X_{\cali}= \{x_k: k \in \cali \}$ as covariate set.   
If it is too hard to find the smallest covariate set $\cala$ satisfying (\ref{model svs}) especially when $p >> n$, a weaker goal is to find a covariate set  
containing $\cala$ with size as small as possible, which is referred to as sufficient variable screening and is the focus of this paper.  

Research on sufficient variable screening in ultrahigh dimensional setting has gained considerable momentum in recent years. \citet{DCSIS2012} and \citet{ShaoZhang2014} proposed to use marginal distance correlation and marginal martingale difference divergence for sufficient variable screening.  Noticing the close relationship between sufficient variable selection and sufficient dimension reduction (\citealp{SIR1991}; \citealp{Cook:1998}), \citet{Yu2014}, \citet{Yin2015} and \citet{MSIR2016} developed different dimension reduction based screening methods.

In many biomedical studies, the response are often censored rather than fully observed. We consider survival data in which $\T$ is the true lifetime, $\C$ is the censoring time and 
we only observe $\T^o= \min \{\T,\C\}$ and the censoring indicator $\delta=I(\T\leq \C)$. 
Sufficient variable selection with censored response is finding $\cala$ in (\ref{model svs}) with $Y$ replaced by $  
(\T, \C)$, i.e., 
\begin{align}\label{model csvs}
(\T,\C) \indep \X \mid \X_{\cala}. 
\end{align}
While our focus is (\ref{model csvs}), we can only  observe $(\T^o , \delta )$, instead of $(T,C)$.

There exists very limited amount of work on model-free variable screening with censored responses.  Assuming $\T \indep \C \mid \X$, the quantile  adaptive sure independence screening procedure proposed by 
\citet{QaSIS2013} can be naturally extended to survival analysis. \citet{SII2016} proposed a survival impact index, which characterizes the impacts of a covariate on the distribution of true lifetime $\T$ by evaluating the absolute deviation of the covariate-stratified survival distribution from the unstratified survival distribution. The proposed survival impact index based screening seems to take some advantages over quantile adaptive sure independence screening when dealing with censored responses.

We in this paper give a modification of the directional regression (\citealp{DR2007}) capable for sufficient dimension reduction with censored response%Inspired By A Novel Marginal Sliced Inverse Regression index For The Purpose Of Ultrahigh Dimensional Feature Selection Proposed By \Citet{Msir2016},
, and then characterize a suitable modified directional regression index for sufficient variable screening. The sure screening property is established in the ultrahigh dimensional setting, i.e., with probability tending to one, the smallest covariate set $\cala$ is contained in the set of covariates selected by our proposed procedure. %In the simulation, it can be seen that our screening is computational efficient and performs well to deal with ultrahigh dimensional data with censored response.
We also discuss the limitations of such modified directional regression index and propose a refined iterative procedure of our screening approach. To further enhance the stability of variable screening, we follow  \citet{STABLESELECTION2012} and \citet{STABLESIS2011} to  integrate the resampling scheme into our proposal. 
After screening,  the selected covariate set may contain some unrelevant covariates,   but its size  is much smaller than $n$ so that we may apply variable selection or dimension reduction using an existing method to further reduce the size or dimension of the selected covariate set. 
Our approach are examined through simulation studies and an application to the diffuse large-B-cell lymphoma microarray data \citep{DLBCL data}.

%\csection{Modified Directional Regression Index for Variable Screening}
\section{Modified Directional Regression Index}% for Variable Screening}
\label{sec sdr-sis}

To derive an index for covariate screening, we first reveal a relationship between sufficient variable selection and sufficient dimension reduction, another perspective in reducing covariate dimension. As a by-product,  we extend one method in sufficient dimension reduction, the directional regression, to survival data with censoring, which leads to an index for sufficient variable screening.

%\csubsection{Sufficient variable selection and Sufficient dimension reduction}

Sufficient dimension reduction aims to identify a linear function of $\X$ with dimension lower than $p$,  without losing information. To be specific, we seek a $p\times d$ matrix $\B$ with the smallest $d$ such that  
\begin{align}\label{model sdr}
(\T,C) \indep \X \mid \B\trans\X.
\end{align}
The linear space generated by columns of $\B$ is called the central subspace and denoted as $\mathcal{S}_{(\T,\C)\mid \X}$. The following result reveals a deep connection between sufficient variable selection (\ref{model csvs}) and sufficient dimension reduction (\ref{model sdr}) for censored responses. 	
\begin{proposition}\label{theo: svs and sdr}
	Let $\bfbeta_1,\ldots,\bfbeta_d$ be columns of $\B$ in (\ref{model sdr}) and $\e_k$ be the $p\times 1$ vector whose $k$th element is 1 all other elements are 0. 		 Then, 
	$\sum_{j=1}^d|\e_k\trans\bfbeta_j|>0$ for $k \in \cala$ and  $\sum_{j=1}^d|\e_k\trans\bfbeta_j|=0$ for $k \not \in \cala$, where 
	$\cala$ is given in (\ref{model csvs}). 
\end{proposition}

%\textcolor{red}{Rewrite two paragraphs above w.r.t. the changes in (A3): Let $\Z=\bSig^{-1/2}(\X-\bfmu)$ be the standardized covariate, where $\bfmu=E(\X)$ and $\bSig=\var(\X)$. Then, $\mathcal{S}_{(\T,\C)\mid \X}={\bSig^{-1/2}}\mathcal{S}_{(\T,\C)\mid \Z}$. A key result for the success of  the directional regression in  \cite{DR2007} is that, if there is no censored response so that $(T_o,\delta)$ and $(\T,\C)$ in conditions (A1)-(A3) of Proposition (\ref{pro: Censored RD in CS}) are replaced by $T$, then these three conditions guarantee that the column space of $E[ 2\I_p-E\{(\Z-\widetilde{\Z})(\Z-\widetilde{\Z})\trans\mid \T,\widetilde{\T}\}]^2 $ is equal to $\mathcal{S}_{\T\mid\Z}$, where $\I_p$ is the identity matrix of order $p$ and $(\widetilde{\Z},\widetilde{\T})$ is an independent copy of $(\Z,\T)$. However, in survival analysis, two random variables $\T$ and $\C$ are of interest. Moreover we can only observe $(\T^o,\delta)$ instead of $(T,C)$.  The next proposition extends the result in \cite{DR2007} to the space $\mathcal{S}_{(\T,\C)\mid \Z}$.}

This result tells us that $\B$ in sufficient dimension reduction can be also used for sufficient variable selection.  
Inspired by this, in the following we first extend the directional regression \citep{DR2007} to find the central space $\mathcal{S}_{(\T,\C)\mid \X}$ using survival data with censoring. 

Let 
$\Z=\bSig^{-1/2}(\X-\bfmu)$ be the standardized covariate, where 
$\bfmu=E(\X)$ and $\bSig=\var(\X)$. Then, 
$\mathcal{S}_{(\T,\C)\mid \X}={\bSig^{-1/2}}\mathcal{S}_{(\T,\C)\mid \Z}$.   
A key result for the success of  the directional regression in  \cite{DR2007} is that, 
if $(T,C)$ is observed,  the  column space  of $E[ 2\I_p-
E\{(\Z-\widetilde{\Z})(\Z-\widetilde{\Z})\trans\mid \T,\widetilde{\T},\C,\widetilde{\C}\}]^2 $ is equal to   
$\mathcal{S}_{(\T,\C)\mid\Z}$, where $\I_p$ is the identity matrix of order $p$ and 
$(\widetilde{\Z},\widetilde{\T},\widetilde{\C})$ is an independent copy of $(\Z,\T,\C)$.
However, in survival analysis  $(\T,\C)$ is unobservable; instead, we observe  $(\T^o,\delta)$. The next proposition extends the result in \cite{DR2007} to the survival data with censoring. 

\begin{proposition}\label{pro: Censored RD in CS}
	Let $\M=E[2\I_p-E\{(\Z-\widetilde{\Z})(\Z-\widetilde{\Z})\trans\mid \T^o ,\widetilde{\T^o},\delta,\widetilde{\delta}\}]^2$, where 
	$(\widetilde{\Z},\widetilde{\T^o},\widetilde{\delta})$ is an independent copy of $(\Z,\T^o,\delta )$.
	\\
(i) Suppose that 
\begin{itemize}
	\vspace{-8pt}
	\item [(A1)] For any $\bfnu \in \R^p$ and $\bfnu \bot \mathcal{S}_{(\T,\C)\mid\Z}$, $E(\bfnu\trans \Z\mid P\Z)$ is a linear function of $\Z$ for any  projection $P$ onto $\mathcal{S}_{(\T,\C)\mid\Z}$;
	\vspace{-10pt}
	\item [(A2)] For any $\bfnu \in \R^p$ and $\bfnu \bot \mathcal{S}_{(\T,\C)\mid\Z}$, $\var(\bfnu\trans \Z\mid P\Z)$ is nonrandom for any  projection $P$ onto $\mathcal{S}_{(\T,\C)\mid\Z}$.
	\vspace{-8pt}
\end{itemize}
Then column space of $\M$ is contained in $\mathcal{S}_{(\T,\C)\mid \Z}$.\\
(ii) Suppose further that 
\begin{itemize}
	\vspace{-8pt}
	\item[(A3)] For any $\bfpsi \in \mathcal{S}_{(\T,\C)\mid \Z}$, $\bfpsi \neq 0$, the random variable $E\{[\bfpsi\trans(\Z-\widetilde{\Z})]^2\mid \T^o ,\widetilde{\T^o},\delta,\widetilde{\delta}\}$ is  not equal  to a constant almost surely.
	\vspace{-8pt}
\end{itemize}
Then column space of $\M$ is equal to $\mathcal{S}_{(\T,\C)\mid \Z}$. 
\end{proposition}

Conditions (A1) and (A2) are known as linear conditional mean condition and constant conditional variance condition in the the sufficient dimension reduction literature; see \cite{Cook2007} and \cite{DR2007} for more discussions. Condition (A3) is generally considered to be very mild. See \cite{CRSDR2005} for more details.

Proposition \ref{pro: Censored RD in CS} suggests that we can utilize $\M$ for estimating $\mathcal{S}_{(\T,\C)\mid\Z}$. In applications  we use $\G$, a discretized version of $\M$, to estimate $\mathcal{S}_{(\T,\C)\mid \Z}$. We partition the sample space of the uncensored observations with $\delta=1$ into $H_1$ non-overlapping intervals  $I_{11},\ldots,I_{1H_1}$, and the sample space of censoring time $C$ with $\delta=0$ into $H_0$ non-overlapping intervals $I_{01},\ldots,I_{0H_0}$. Let $p_{lj}=E[I(\delta=l,\T^o \in I_{lj})]$ and $\D_{ijlm}=E[(\Z-\widetilde{\Z})(\Z-\widetilde{\Z})\trans\mid \delta=i,\T^o \in I_{ij},\widetilde\delta=l,\widetilde{\T^o} \in I_{lm})]$, where $(i,j,l,m)\in\{i,j,l,m: i, l=0~\text{or}~1,~j=1,\dots, H_i,~m=1,\dots, H_l\}$. Then $\G$ is expressed as follows
\begin{align}\label{M-discretized}
\G
%&=2\sum[E^2(\Z\Z\trans-I_p\mid \delta=l,\Y\in I_{lj})p_{l,j}]\notag\\
%&~~~+2\{\sum[E(\Z\mid \delta=l,\Y\in I_{lj})E(\Z\trans\mid \delta=l,\Y\in I_{lj})]p_{l,j}\}^2\notag\\
%&~~~+2\sum[E(\Z\trans\mid \delta=l,\Y\in I_{lj})E(\Z\mid \delta=l,\Y\in I_{lj})p_{l,j}]\notag\\
%&~~~\times\sum[E(\Z\mid \delta=l,\Y\in I_{lj})E(\Z\trans\mid \delta=l,\Y\in I_{lj})p_{l,j}]\notag\\
%&=\sum_{lj} 2p_{l,j}(p_{l,j}^{-1}\V_{l,j}-I_p)^2+2(\sum_{lj} \U_{l,j}\U_{l,j}\trans p_{l,j}^{-1})^2+2(\sum_{lj} \U_{l,j}\trans \U_{l,j}p_{l,j}^{-1})(\sum_{lj} \U_{l,j}\U_{l,j}\trans p_{l,j}^{-1})\notag
&=\sum_{ij}\sum_{lm}p_{ij}p_{lm}(2\I_p-\D_{ijlm})^2.
\end{align}
We can recover $\mathcal{S}_{(\T,\C)\mid \Z}$ through the  eigen-decomposition
$	\G\bfeta_i=\lambda_i \bfeta_i$, where $\lambda_i$'s are scalars and $\bfeta_i$'s are $p \times 1$ vectors, and obtain  $\B =(\bSig^{-1/2}\bfeta_1,\ldots,\bSig^{-1/2}\bfeta_d)$.

As $\bSig^{-1/2}$ is involved in $\M$, $\G$, and $\B$, the classical sufficient dimension reduction methods fail to work when $p>n$ unless we have a good estimator of  $\bSig^{-1/2}$. 

For sufficient variable selection, we do not need the entire matrix $\G$ in (\ref{M-discretized}). Proposition \ref{prop:gkstar} below shows that the following marginal utility of $\G$, 
\begin{align}\label{def:gkstar}
g_k^*=\sum_{ij}\sum_{lm}p_{i,j}p_{l,m}[\e_k\trans{\bSig}^{-1/2}(2\I_p-\D_{ijlm}){\bSig}^{-1/2}\e_k]^2,
\end{align}
is a perfect index for sufficient variable selection.  Note that $\e_k\trans{\bSig}^{-1/2}(2\I_p-\D_{ijlm}){\bSig}^{-1/2}\e_k$ is not the $k$th diagonal element of the matrix $2\I_p-\D_{ijlm}$
in (\ref{M-discretized}), but the $k$th diagonal element of the matrix ${\bSig}^{-1/2}(2\I_p-\D_{ijlm}){\bSig}^{-1/2}$.

\begin{proposition}\label{prop:gkstar}
	If conditions  (A1)-(A3) hold, then $g_k^*>0$ if $k\in \cala$ and $g_k^*=0$ if $k\not \in \cala$.
\end{proposition}

The next result gives an alternative expression of $g_k^*$, which is useful for our derivation.

\begin{lemma}\label{Lemma: rewriten gkstar}
	Let $\U_{lj}=E[\Z I(\delta=l,\T^o \in I_{lj})]$ and $\V_{lj}=E[\Z\Z\trans I(\delta=l,\T^o \in I_{lj})]$.
	Then	
	\begin{align}\label{def:gkstar1}
	g_k^*=2\sum_{lj} p_{lj}[\e_k\trans{\bSig}^{-1/2}(p_{lj}^{-1}\V_{lj}-\I_p){\bSig}^{-1/2}\e_k]^2+4\left(\sum_{lj} p_{lj}^{-1}\e_k\trans {\bSig}^{-1/2}\U_{lj}\U_{lj}\trans{\bSig}^{-1/2}\e_k\right)^2
	\end{align}
\end{lemma}

However, $g_k^*$ in (\ref{def:gkstar}) or (\ref{def:gkstar1}) still involves ${\bSig}^{-1/2}$ which is hard to estimate when $p$ is bigger than or comparable to $n$. 
We then follow the idea in  independence variable screening (\citealp{SIS2008}, \citealp{DCSIS2012}, and \citealp{MSIR2016}), i.e., we replace  ${\bSig}^{-1/2}$ in 
(\ref{def:gkstar1}) by $\I_p$ and obtain the following  modified directional regression index, 
\begin{align}\label{def:gk}
g_k=2\sum_{lj} p_{lj}(V_{ljk}/p_{lj}-1)^2+4\left(\sum_{lj} U_{ljk}^2/p_{lj}\right)^2
\end{align}
where 
$U_{ljk}=E[z_kI(\delta=l,\T^o \in I_{lj})]$ and $V_{ljk}=E[z_k^2 I(\delta=l,\T^o \in I_{lj})]$, 
$z_k=(x_k-\mu_k)/\sigma_k$, $\mu_k=E(x_k)$, and $\sigma_k^2={\rm Var}(x_k)$.
Although $g_k$ in (\ref{def:gk}) is not a prefect index for sufficient variable selection as ${\bSig}^{-1/2}$ may be incorrectly treated as $\I_p$, it is good enough for sufficient variable screening, i.e., finding a set containing $\cala$ in (\ref{model csvs})
under some conditions. The following result is an example, in which the conditions 
are similar to those in  \cite{gk assumption 2015} and \cite{MSIR2016}.

\begin{proposition}\label{prop1_m1}
	Assume conditions (A1)-(A3). Suppose also that  ${\rm Cov}(x_i,x_j)$ has the same sign for $i,j\in\cala$, and that there exists $h\in\{1,\dots,d\}$ such that the $(j,h)$th element of $\B$ in (\ref{model sdr}) 
	have the same sign for all $j\in\cala$. Then $g_k>0$ if $k\in \cala$.
	%	\item[\textup{(2).}] Assume $Cov(\X_{\cala},\X_{\cala^c})=\0$. Then $g_k=0$ if $k\in \cala^c$.	
\end{proposition}	
%\begin{proposition}\label{prop1_m2}
%	In addition to (A1), (A2) and (A3), we assume $Cov(\X_{\cala},\X_{\cala^c})=\0$. Then $g_k=0$ if $k\in %\cala^c$.	
%\end{proposition}

%\begin{align*}
%g_k^*&=2\sum_{lj} p_{l,j}[\frac{\e_k\trans{\bSig}^{-1/2}(\V_{l,j}-\I_p){\bSig}^{-1/2}\e_k}{p_{l,j}}]^2+4\left(\sum_{lj}\frac{\e_k\trans {\bSig}^{-1/2}\U_{l,j}\U_{l,j}\trans{\bSig}^{-1/2}\e_k}{p_{l,j}}\right)^2\\
%g_k&=2\sum_{lj} p_{l,j}[\frac{\e_k\trans({O}^{-1/2}{\bSig}^{1/2}\V_{l,j}{\bSig}^{1/2}{O}^{-1/2}-\I_p)\e_k}{p_{l,j}}]^2+4\left(\sum_{lj}\frac{\e_k\trans {O}^{-1/2}{\bSig}^{1/2}\U_{l,j}\U_{l,j}\trans{\bSig}^{1/2}{O}^{-1/2}\e_k}{p_{l,j}}\right)^2\\
%  &=2\sum_{lj} p_{l,j}[\frac{\e_k\trans({O}^{-1/2}{\bSig}^{1/2}(\V_{l,j}-\I_p){\bSig}^{1/2}{O}^{-1/2}\e_k}{p_{l,j}}]^2+4\left(\sum_{lj}\frac{\e_k\trans {O}^{-1/2}{\bSig}^{1/2}\U_{l,j}\U_{l,j}\trans{\bSig}^{1/2}{O}^{-1/2}\e_k}{p_{l,j}}\right)^2\\
%  &=2\sum_{lj} p_{l,j}[\frac{\e_k\trans({O}^{-1}{\bSig}^{1/2}\V_{l,j}{\bSig}^{1/2}-\I_p)\e_k}{p_{l,j}}]^2+4\left(\sum_{lj}\frac{\e_k\trans {O}^{-1}{\bSig}^{1/2}\U_{l,j}\U_{l,j}\trans{\bSig}^{1/2}\e_k}{p_{l,j}}\right)^2\\
%O&=\sum_{i=1}^p \e_i\e_i\trans{\bSig}\e_i\e_i\trans=\diag(\sigma_1^2,\dots,\sigma_p^2)\\
%p_{l,j}&=E[I(\delta=l,\Y\in I_{lj})]\\
%\U_{l,j}&=E[\Z I(\delta=l,\Y\in I_{lj})]\\
%\V_{l,j}&=E[\Z\Z\trans I(\delta=l,\Y\in I_{lj})]\\
%\end{align*}

%\csection{Sure Independence Screening}
\section{Sure Independence Screening}

In this section we show that variable screening by using the index $g_k$ in (\ref{def:gk})
holds some asymptotic properties under some conditions. Procedures with weaker conditions are considered in the next section. 
Let $(x_{ki}, t^o_i, \delta_i)$, $i=1,...,n$, $ k=1,...,p$, be observations from the random sample from $(\X, \T^o, \delta )$,  
$\hat \mu_k = \sum_{i=1}^n x_{k,i}/n$, $\hat\sigma_k=\{\sum_{i=1}^n(x_{k,i}-\hat\mu_k)^2/n\}^{1/2}$, 
$\widehat{z}_{ki} = (x_{ki}- \hat\mu_k)/\hat\sigma_k$,
$\widehat p_{lj}=\sum_{i=1}^n I(\delta_i=l,t^o_i  \in I_{lj})/n$, $\widehat U_{ljk}=\sum_{i=1}^n \widehat{z}_{ki} I(\delta_i=l,t^o_i  \in I_{lj})/n$, and $\widehat V_{ljk}=\sum_{i=1}^n \widehat{z}_{ki}^2  I(\delta_i=l,t^o_i  \in I_{lj})/n$. A sample estimator of $g_k$ in (\ref{def:gk}) is $\widehat g_{k}$ defined by (\ref{def:gk}) with $p_{lj}$, $U_{ljk}$, and $V_{ljk}$ replaced by 
$\widehat p_{lj}$, $\widehat U_{ljk}$, and $\widehat V_{ljk}$, respectively. 
We select the set of covariates such that $\widehat g_k$ is large enough. Define
\begin{align}\label{screening}
\widehat \cala=\{k: \widehat g_k \ge \gamma, 1\le k \le p\},
\end{align}
where $\gamma$ is a threshold to be specified later. 
To study the theoretical property of $	\widehat \cala$ in 
(\ref{screening}), we consider the following conditions:
\begin{itemize}
	\vspace{-6pt}
	\item [(C1)] $p>n$ and $\log p=O(n^\xi)$ for some $\xi \in (0,1-2\kappa)$, where $\kappa$ is given in condition (C3);
	\vspace{-12pt}
	\item [(C2)] There exist some
	$0< \varsigma <1/4$ such that  $E\{\exp(tz_k^2)\}\le K_0$ for $1\le k \le p$ and all $|t|\le \varsigma$, where $K_0$ is a fixed constant;
	\vspace{-12pt}
	\item [(C3)] $\min_{k\in\cala} g_k>2c_0 n^{-\kappa}$ for some constants $c_0>0$ and $0 \le \kappa\le 1/2.$
	\vspace{-6pt}
\end{itemize}
Condition (C1) was also used by \citet{SIS2008} and \cite{RANKSIS2012}, which allows $p$ to be as large as an exponential of the sample size $n$. Condition (C2) assumes that  all covariates have an exponential-type tails, which is a common technique condition in ultrahigh dimensional data analysis; see, for example, \citet{CAI2011}.
Condition (C3) is naturally motivated from Proposition \ref{prop1_m1}, and requires that the index $g_k$ for $k\in \cala$ is not too small, which is also a common condition in the literature of sure independence screening \citep{SIS2008,RANKSIS2012,DCSIS2012}.

The next theorem confirms the sure screening property of  $\widehat \cala$.

\begin{theorem}\label{theo: sure screening}
	(i) Assume conditions (C1) and (C2). Then
	\begin{align*}
	\p\left\{\max_{1\le k\le p} |\widehat g_k-g_k|\ge C_0(\log p/n)^{1/2}\right\}\le 72p^{-\tau-1},
	\end{align*}
	where $\tau>0$ is a constant and $C_0$ is defined in (\ref{def: C0}) in the Appendix.\\
	(ii)		 Additionally, if condition (C3) also holds and $\gamma\le c_0 n^{-\kappa}$, then 
	\begin{align}\label{prob}
	\p\left\{\cala\subseteq \widehat\cala\right\} \ge 1-72p^{-\tau-1}, 
	\end{align}
	where $\widehat \cala$ is given by (\ref{screening}).
\end{theorem}

Since $g_k$  in (\ref{def:gk}) is a  modified directional regression index and 
Theorem \ref{theo: sure screening} indicates that the probability in (\ref{prob}) converges to one as $n$  diverges to infinity,  
we name the proposed covariate screening procedure as the modified directional regression-sure independence screening (\MDRSIS) method.
Note that ${\bSig}^{-1/2}= \I_p$ is assumed in the derivation of $g_k$, but it is not needed in establishing the result in Theorem \ref{theo: sure screening}, as long as (C1)-(C3) hold true. In the next section we obtain some further results in the case where (C3) may be violated.

The threshold value $\gamma$ depends on constants $c_0$ and $\kappa$ in (C3), which is unknown in real applications.
We follow the convention developed in \citet{SIS2008} 
and define the  screened covariate set as 
\begin{equation}\label{calas}
\widehat \cala^*=\{k :
\widehat g_k\geq \widehat g_{d_n} \},
\end{equation}
where
$\widehat  g_{d_n}$ is the $d_n$th largest ranked index among all
$\widehat g_k$'s. Following \cite{SIS2008}, $d_n$ can be set as $\lfloor n/\log n\rfloor$, where $\lfloor a \rfloor$ denotes the integer part of $a$. Theorem \ref{theo: sure screening} together with Theorem 1 in \cite{SIS2008} guarantee $\p( \cala \subseteq \widehat \cala^* )$ converges to  one as $n\to \infty$.

Let $\X_{\calb} = \{\x_k : k \in \calb\}$ be the smallest covariate set related to  the life time  $\T$, i.e., $\calb$ satisfies  $\T \indep \X\mid \X_{\calb}$. 
Sometimes we are interested in identifying $\calb$  instead of $\cala$.
For example, if  we assume $\T \indep \C\mid \X$, which is typically needed for 
many survival analysis methods although it is not needed  for the asymptotic property of \MDRSIS, then  $\T \indep \C\mid \X$ and  $\T \indep \X\mid \X_{\calb}$
imply $\T \indep \C\mid \X_{\calb}$ so that survival analysis can be carried out using $\X_\calb$. However, identifying $\cala$ may result in a more efficient analysis if information on $C \mid \X$ is useful. 

Since $\calb \subseteq \cala$, the sure screening property $\p( \calb \subseteq \widehat \cala^* )\rightarrow 1$ can still be achieved based on Theorem \ref{theo: sure screening}. 
Unless $\C \indep \X\mid \X_{\calb}$,  $\calb$ is a strict subset of $  \cala$.
Even If we focus on $\calb$ only, it is unnecessary to do covariate screening to find 
a $\widehat \calb^*$ with $\p( \calb \subseteq \widehat \calb^* )\rightarrow 1$, because both $\widehat \calb^*$ and $\widehat \cala^*$ are screening methods aimed to reduce the size of  covariate set to a manageable number $< n$ and a further dimension reduction or variable selection can be applied to $\widehat \cala^*$ 
%Hence, we recommend to apply another dimension reduction to the covariates in $\widehat \cala^*$. 
as the size of $\widehat \cala^*$ is much smaller than $n$, 
i.e., $d_n /n \to 0$.
% there exist sufficient dimension reduction methods with good asymptotic properties. 
%If we aim for  $\calb$, then we can apply the method in \cite{DRcensor1999}. If $\cala$ is considered, then we can use the method we described in Section 2, which is an extension of the method in  \citet{DR2007} to censored data. Specifically, we can use $\G$ defined in (\ref {M-discretized}) with $\bSig$ estimated by the sample covariance matrix of the covariate vector $\X_{\widehat \cala^*}$ with dimension $d_n <n$. 

%\csection{Enhanced Screening with Iteration and Resampling}
%\csubsection{Iterative variable screening}

\section{Enhanced Screening with Iteration and Resampling}
\subsection{Iterative variable screening}
Condition (C3) plays a key role for the sure independence screening property of \MDRSIS. 
However, (C3) may be violated since $g_k$ ignores information contained in $\bSig$. The next result identifies a situation where (C3) does not hold.

\begin{proposition}\label{propgk}
	Let $\bfbeta_1,\ldots,\bfbeta_d$ be columns of $\B$ in (\ref{model sdr}). For any $\bSig$, if there exists $k \in \cala$ such that  $\sum_{i=1}^d|\e_k\trans{\bSig}\bfbeta_i|=0$, then $g_k=0$ and, hence, (C3) is violated. 
\end{proposition}

In the situation described by Proposition \ref{propgk}, the sure screening property can not be guaranteed. To circumvent this issue, we should handle the correlations among covariates and consider iterative screening. Suppose that we have already selected a   covariate set  $\X_\calf=\{\x_k:
k\in \calf\}$, where  $\calf \subset \{ 1,...,p \}$. 
Define  $\bfmu_{\calf}=E(\X_\calf)$ and ${\bSig}_\calf=\var(\X_\calf)$.  For any  $e
\not\in \calf$, let ${\bSig}_{\calf,e}={\cov}({\X}_{\calf},x_e)$ and  ${\x}_{e\mid\calf}={\x}_e-{\bSig}_{{\calf},e}\trans{\bSig}_\calf^{-1}\X_{\calf}$ be the residual of $\x_e$ regressed on $\X_{\calf}$. Then,  $\cov(\x_{e\mid\calf},\X_{\calf})=0$, which suggests that we can adopt  
%\textcolor{red}{"(an iterative independence modified directional regression index similar with $g_k$ based on $(\x_{e\mid\calf},\T^o,\delta)$)"}
the marginal utility of modified directional regression based on $(\x_{e\mid\calf},\T^o,\delta)$ as an index for iterative screening. Define $\mu_{e\mid\calf}=\mu_e-{\bSig}_{\calf,e}\trans{\bSig}_\calf^{-1}\bfmu_{\calf}$, ${\sigma}^2_{e\mid\calf}={\sigma}^2_e-{\bSig}_{\calf,e}\trans{\bSig}_\calf^{-1}{\bSig}_{\calf,e}$, and $z_{e\mid\calf}=(\x_{e\mid\calf}-\mu_{e\mid\calf})/{\sigma}_{e\mid\calf}$ as the standardized version of $\x_{e\mid\calf}$. Then we define the following iterative modified directional regression index:
\begin{align*}
g_{e\mid\calf}=2\sum_{lj} p_{lj}(V_{lje\mid\calf}/p_{lj}-1)^2+4\left(\sum_{lj} U_{lje\mid\calf}^2/p_{lj}\right) ^2, \quad \quad e\not \in \calf ,
\end{align*}
where %$p_{l,j}=E[I(\delta=l,\Y\in I_{lj})]$, 
$U_{lje\mid\calf}=E[z_{e\mid\calf}I(\delta=l,\T^o \in I_{lj})]$ and $V_{lje\mid\calf}=E[z_{e\mid\calf}^2 I(\delta=l, \T^o \in I_{lj})]$. The next proposition illustrates the advantage of the proposed iterative screening method.

\begin{proposition}\label{propgki}
	Let $\calf$ be a nonempty subset of 
	$ \{1,...,p\}$. Suppose that 
	\begin{itemize}
		\vspace{-6pt}
		\item[(C4)] $\min_{k\in \cala,i=1,\ldots,d}|\beta_{ik}|>c_1n^{-\theta}$ for some constants $c_1>0$ and $0< \theta\le 1/8$, where  $\beta_{ik}$ is the $(i,k)$th element of 
		$\B$ in (\ref{model sdr});
		\vspace{-6pt}
		\item[(C5)] $\sigma^2_{e\mid\calf}\geq c_2$ for some constant $c_2>0$, where $e \not \in \calf$.
		\vspace{-8pt}
	\end{itemize}
	Then $g_{e\mid\calf}>2c_0n^{-\kappa}$ for $e\in \cala$ with some constants $c_0>0$ and $0 \leq \kappa\le 1/2$.
\end{proposition}
Condition (C4) is a mild condition previously used by \citet{SIS2008}. Condition (C5) means that the $e$th relevant covariate missed in the previous steps should not be expressed only by the set of covariates selected by previous steps, which is a general condition under iteration construction.   

The result of this proposition illustrates that utilizing the index $g_{e\mid\calf}$ is able to identify the informative predictors missed by \MDRSIS. To illustrate, suppose that 
$\widehat\cala^*_1 =\widehat\cala^*$ as define by (\ref{calas}) is selected by \MDRSIS. Suppose that we carry out one iteration to obtain 
a covariate set $\widehat\cala^*_2 =\{e: e \not \in \widehat\cala^*_{1}, 
\widehat g_{e\mid\widehat\cala^*_{1}} \geq \widehat g_{e\mid\widehat\cala^*_{1},q}\}$, where $\widehat g_{e\mid\widehat\cala^*_{(1)},q}$ is the $q$th largest ranked index among all $\widehat g_{e\mid\widehat\cala^*_{1}}$'s.
By Proposition \ref{propgk},  $\widehat\cala^*_{2}$ may recover some relevant covariates missed by  $\widehat\cala^*_{1}$ selected by \MDRSIS, with an appropriate choice of $q$. The covariate set after iteration is 
$ \widehat\cala^*_{1} \cup \widehat\cala^*_{2}$. 
Numerical studies show that $q$ can be much smaller than $\lfloor n/\log n\rfloor$. 

Although $ \widehat\cala^*_{1} \cup \widehat\cala^*_{2}$ is better than 
$ \widehat\cala^*_{1} = \widehat\cala^*$ in terms of containing relevant covariates,  
its size is always larger than the size of $ \widehat\cala^*$. Hence, to apply the iterative variable screening, we do not have to start with  $ \widehat\cala^*_{1} = \widehat\cala^*$, especially when we doubt about whether \MDRSIS \
can select all relevant covariates. 
Instead, we may start with  a $ \widehat\cala^*_{1}$ smaller than 
$ \widehat\cala^*$ and set the size of final covariate sets selected after iterations
to be the same as that of $ \widehat\cala^*$. This leads to the following general iterative procedure for covariate screening. 
\begin{description}
	\vspace{-8pt}
	\item Step 1. Based on  $\widehat g_{k}$ $(k=1,\dots,p)$, we select $p_{1}$ covariates by \MDRSIS. Denote the set of indices of selected covariates by $\widehat\cala^*_{1}$.
	\vspace{-10pt}
	\item Step 2.  For $e \not \in \widehat\cala^*_{1}$, we estimate $g_{e\mid\widehat\cala^*_{1}}$ by a sample estimator $\widehat g_{e\mid\widehat\cala^*_{1}}$. Based on  $\widehat g_{e\mid\widehat\cala^*_{1}}$, we select $p_{2}$ covariates by \MDRSIS \ with the resulting covariate set denoted by $\widehat\cala^*_{2}$.
	\vspace{-10pt}
	\item Step 3. Repeat Step 2  until the total selected
	number of covariates is $d_n$. The final selected covariate set is then $\widehat \cala_{I}=\widehat \cala^*_{1}\cup\cdots\cup\widehat \cala^*_{S}$, where $\sum_{v=1}^S p_v = d_n$. 
\end{description}
We name 
this iterative procedure as the modified directional regression-iterative sure independent screening  (\MDRIS) method. 
Under conditions (C1)-(C2) and (C4)-(C5), it can be shown similarly to Theorem 3.1 that 
$\p ( \cala \subset \widehat \cala_I) \to 1$ as $n \to \infty$. 
Some simulation results are presented in Section 5 for the selection of $p_v$'s and the results show that $S=2$ works well under our simulated models.

\subsection{Stability Screening}
While \MDRIS \ is used to improve \MDRSIS \ in including all relevant covariates, 
the stability selection approach introduced in 
\citet{STABLESELECTION2012} 
is designed to  reduce the number of falsely selected covariates 
through combining resampling with high dimensional variable selection.
\citet{STABLESIS2011}  adapted this resampling mechanism to iterative sure independence screening for genome-wide association studies. Along with their
developments, we further propose the following procedure to improve \MDRIS. 
The algorithm is based on $B$ independent subsamples of size $n_s< n$ without replacement from the training data set. 
For the $b$th subsample, we apply \MDRIS \ to select a candidate covariate set $\widehat \cala_{I}^{(b)}$.
The stability screened covariate set based on this  procedure is
\begin{align*}
\widehat \cala_{S}=\{k: \pi_k\ge \pi_0\}, \quad\quad 
\pi_k=\frac{1}{B} \sum_{i=1}^B I(k\in \widehat\cala_{I}^{(i)}), 
\end{align*}
where $I( \cdot )$ is the indicator function. Following \citet{STABLESIS2011}, we prespecify threshold value $\pi_0$ to be 0.3 or 0.4 in practical use.
We name this procedure as the modified directional regression-stability sure independence screening  (\MDRSS) method. In Section 5, we compare \MDRSS \ with \MDRIS \ in simulations. 

%\csection{Numerical Results}
\section{Numerical Results}
In this section, we assess the performance of the proposed \MDRSIS, \MDRIS \ and \MDRSS \ by Monte Carlo simulation. We further examine the proposed screening procedure with an empirical analysis of a real-data example. 
 
%\csubsection{Simulation study}
\subsection{Simulation study} 
The covariate vector $\X$ is generated from the multivariate normal distribution with mean $0$ and  covariance matrix $\sig$ whose $(i,j)$th element is $\rho^{|i-j|}$ with  $\rho= 0$, $0.4$, or $0.8$ throughout our simulations. Let $\epsilon\sim N(0,1)$ be an error term  independent of $\X$ and the censoring time $C$.  We consider the following five models representing various types of covariate functions with different degree of nonlinearity, and multiple failure and censoring distributions. 
 \begin{description}
 	\item
 	M1. $\T=(2\X\trans\bfbeta_1)^2+12\sin(3\X\trans\bfbeta_2/7)+0.2\epsilon$, $\C \sim N(0, 4) - N(5, 1) + N(15, 1)$, where $\bfbeta_1$ and  $\bfbeta_2$ are $p\times 1$ vectors with their first six components being $(1,0,1,0,0,0)\trans$ and $(0,0,0,0,1,1)\trans$, respectively, and rest components being zeros. 
 	\item
 	M2. $\T=(2\X\trans\bfbeta_1)^2+|8\X\trans\bfbeta_2|+0.2\epsilon$, $\C \sim N(0, 4)- N(5, 1) + N(30, 1)$, where $\bfbeta_1$ and  $\bfbeta_2$ are same as those in (M1). 
 	\item
 	M3. $\T=10\sin(\X\trans\bfbeta_1/4)+4|\X\bfbeta_2\trans|+0.2\epsilon$, $\C \sim N(0, 4)- N(5, 1) + N(15, 1)$, where $\bfbeta_1$ and  $\bfbeta_2$ are same as those in (M1).
 	\item 
 	M4. $\T=\exp(\X\trans\bfbeta_1)+|(\X\bfbeta_2\trans)^3|+0.2\epsilon$, $\C \sim N(0, 4)- N(5, 1) + 4N(30, 1)$, where $\bfbeta_1$ and  $\bfbeta_2$ are $p\times 1$ vectors with their first six components being $(-4,4,3,0,0,0) \trans$ and $(0,0,1,0,1,0)\trans$, respectively, and rest components being zeros.
 	\item 
 	M5. $\T=1.5(\X\trans\bfbeta_1)^2+\exp(\X\bfbeta_2\trans)+0.2\epsilon$, $\C=\X\trans\bfbeta_3+8$, where $\bfbeta_1$, $\bfbeta_2$ and  $\bfbeta_3$ are $p\times 1$ vectors with their first six components being $(1,0,0,0,0,0)\trans$, $(1,2,2,0,0,0)\trans$ and $(0,0,1,0,0,1)\trans$, and rest components being zeros.
 \end{description}
 In all models, $T \indep C \mid \X$. 
 In M1-M3, 4 relevant covariates for $T$ are $\x_1,\x_3,\x_{5}$, and $\x_{6}$. In M4, 4 relevant covariates for $T$ are $\x_1,\x_2,\x_{3}$, and $\x_{5}$. In M5, 3 relevant covariates for $T$ are $\x_1,\x_2$, and $\x_{3}$, and 2 relevant covariates for $C$ are $\x_{3}$ and $\x_{6}$. 
 
We first fixed the sample size $n$ to be $200$ and the dimension $p$ to be $400$, and compare our method \MDRSIS \  with SII \citep{SII2016} and QaSIS \citep{QaSIS2013}. To evaluate the performance of the 3  methods, we ran 500 simulations and, for each of the 3 methods, we computed the proportion that an individual relevant predictor was selected and the proportion that all relevant predictors were selected. The simulation results reported in 
Table~\ref{Tab:P} show that our method is the best among all methods in most cases. And for all the cases in which the other two methods perform well, our method performs at least better than the other two methods. 
 
Table~\ref{Tab:T} reports the average computing time of three method with $p=200$ and different values of $n$, or $n=200$ and various values of $p$. 
The computations are performed using R on ECNU IBM Platform Application Center 9.1.3. We can see that our method is the most computational efficient among the three methods and is increasingly more efficient as $n$ and $p$ are larger. Also, SII is computational intensive, which may lead problems in applications with large $n$ and/or $p$. 
  
\begin{table}
	\def~{\hphantom{0}}
	\caption{Simulation proportions of each  relevant covariate and all relevant covariates  selected by  \MDRSIS, SII, and QaSIS with $\alpha =0.5$;  $p=400$, $n=200$, $d_n = \lfloor n/ \log n\rfloor =37$, simulation replication $500$}\label{Tab:P}
	\scalebox{0.78}{
		\begin{tabular}{ccccccccccccccccccc}%{|lr|rrrrr|rrrrrr|rrrrrr|c}
			\hline
			&& \multicolumn{5}{c}{\centering \MDRSIS}&& \multicolumn{5}{c}{\centering SII}&  &\multicolumn{5}{c}{\centering QaSIS}\\
			\cline{3-7} \cline{9-13} \cline{15-19}%\hline			
			&& \multicolumn{4}{c}{\centering relevant covariate}&&&\multicolumn{4}{c}{\centering relevant covariate}&&&\multicolumn{4}{c}{\centering relevant covariate}&\\
			%&\quad& \multicolumn{4}{c}{\centering $\calP_s$}&$\calP_a$&\quad&\multicolumn{4}{c}{\centering $\calP_s$}&$\calP_a$& \quad&\multicolumn{4}{c}{\centering $\calP_s$}&$\calP_a$\\
			model&$\rho$&1&2&3&4& all&&1&2&3&4&all&&1&2&3&4&all\\
			%&$\x_{(1)}$&$\x_{(2)}$&$\x_{(3)}$&$\x_{(4)}$&All&\quad&$\x_{(1)}$&$\x_{(2)}$&$\x_{(3)}$&$\x_{(4)}$&All&\quad&$\x_{(1)}$&$\x_{(2)}$&$\x_{(3)}$&$\x_{(4)}$&All\\
			\hline
			M1& \ \ 0 & 0$\.$90 & 0$\.$95 & 0$\.$91 & 0$\.$86 & 0$\.$66 &   & 0$\.$50 & 0$\.$51 & 1$\.$00 & 1$\.$00 & 0$\.$21 &   & 0$\.$39 & 0$\.$34 & 0$\.$81 & 0$\.$83 & 0$\.$08 \\
			&0.4    & 0$\.$98 & 0$\.$97 & 0$\.$98 & 1$\.$00 & 0$\.$93 &   & 0$\.$70 & 0$\.$80 & 1$\.$00 & 1$\.$00 & 0$\.$57 &   & 0$\.$42 & 0$\.$41 & 0$\.$92 & 0$\.$92 & 0$\.$14 \\
			&0.8    & 1$\.$00 & 1$\.$00 & 1$\.$00 & 1$\.$00 & 1$\.$00 &   & 1$\.$00 & 1$\.$00 & 1$\.$00 & 1$\.$00 & 1$\.$00 &   & 0$\.$80 & 0$\.$83 & 0$\.$90 & 0$\.$90 & 0$\.$57 \\
			M2& \ \ 0 & 0$\.$96 & 0$\.$96 & 0$\.$89 & 0$\.$88 & 0$\.$71 &   & 0$\.$70 & 0$\.$69 & 0$\.$63 & 0$\.$57 & 0$\.$17 &   & 0$\.$51 & 0$\.$49 & 0$\.$63 & 0$\.$63 & 0$\.$11 \\
			&0.4    & 1$\.$00 & 1$\.$00 & 1$\.$00 & 1$\.$00 & 1$\.$00 &\quad& 0$\.$80 & 0$\.$83 & 0$\.$97 & 0$\.$96 & 0$\.$61 &\quad& 0$\.$52 & 0$\.$51 & 0$\.$81 & 0$\.$79 & 0$\.$15\\
			&0.8     & 1$\.$00 & 1$\.$00 & 1$\.$00 & 1$\.$00 & 1$\.$00 &\quad& 1$\.$00 & 1$\.$00 & 1$\.$00 & 1$\.$00 & 1$\.$00 &\quad& 0$\.$82 & 0$\.$93 & 0$\.$98 & 0$\.$95 & 0$\.$74\\
			M3&\ \ 0  & 0$\.$90 & 0$\.$88 & 0$\.$88 & 0$\.$89 & 0$\.$62 &   & 1$\.$00 & 1$\.$00 & 0$\.$60 & 0$\.$60 & 0$\.$36 &   & 0$\.$98 & 1$\.$00 & 0$\.$68 & 0$\.$70 & 0$\.$44 \\
			&0.4    & 0$\.$91 & 0$\.$88 & 1$\.$00 & 1$\.$00 & 0$\.$80 &   & 1$\.$00 & 1$\.$00 & 0$\.$98 & 0$\.$95 & 0$\.$94 &   & 0$\.$98 & 0$\.$97 & 0$\.$92 & 0$\.$91 & 0$\.$80 \\
			&0.8    & 0$\.$99 & 1$\.$00 & 1$\.$00 & 1$\.$00 & 0$\.$99 &   & 1$\.$00 & 1$\.$00 & 1$\.$00 & 1$\.$00 & 1$\.$00&   & 0$\.$96 & 0$\.$97 & 1$\.$00 & 0$\.$99 & 0$\.$93 \\
			M4&\ \ 0  & 0$\.$97 & 0$\.$97 & 0$\.$87 & 0$\.$77 & 0$\.$63 &   & 1$\.$00 & 1$\.$00 & 1$\.$00 & 0$\.$10 & 0$\.$10 &   & 0$\.$58 & 0$\.$59 & 0$\.$81 & 0$\.$87 & 0$\.$26 \\
			&0.4    & 0$\.$25 & 0$\.$88 & 1$\.$00 & 0$\.$95 & 0$\.$19 &   & 0$\.$96 & 1$\.$00 & 1$\.$00 & 0$\.$29 & 0$\.$28 &   & 0$\.$33 & 0$\.$51 & 0$\.$96 & 0$\.$98 & 0$\.$16 \\
			&0.8    & 0$\.$52 & 1$\.$00 & 1$\.$00 & 1$\.$00 & 0$\.$52 &   & 0$\.$57 & 1$\.$00 & 1$\.$00 & 1$\.$00 & 0$\.$57 &   & 0$\.$66 & 0$\.$97 & 1$\.$00 & 1$\.$00 & 0$\.$65 \\
			M5&\ \ 0  & 1.00 & 0.99 & 1.000 & 0.99 & 0.98 &   & 1.00 & 1.00 & 1.00 & 0.22 & 0.22 &   & 0.96 & 0.53 & 0.91 & 0.50 & 0.25 \\
			&0.4    & 1.00 & 1.00 & 1.000 & 1.00 & 1.00 &   & 1.00 & 1.00 & 1.00 & 0.15 & 0.15 &   & 0.96 & 0.68 & 0.85 & 0.53 & 0.33 \\
			&0.8    & 1.00 & 1.00 & 1.000 & 1.00 & 1.00 &   & 1.00 & 1.00 & 1.00 & 0.88 & 0.88 &   & 0.84 & 0.93 & 0.91 & 0.53 & 0.43 \\
			\hline
		\end{tabular}}
	\end{table}
	
\begin{table}
	\centering
	\caption{Computing time (in seconds) required by \MDRSIS, SII and QaSIS with $\alpha=0\.5$ under model M1}\label{Tab:T}	
	\vspace{2mm}
	\scalebox{0.78}{
		\begin{tabular}{cccc ccc cccc}
			\hline
			\multicolumn{4}{c}{$p=400$}&&&&\multicolumn{4}{c}{$n=200$} \\
			$n$& \MDRSIS&QaSIS& SII&&&&$p$ & \MDRSIS&QaSIS& SII\\
			\cline{1-4} \cline{8-11}
			\	\ 200 & 0$\.$17  & \ 0$\.$74 & \ 577$\.$11&&&& \ \ 200 & 0$\.$08   & \ 0$\.$38 & \ \ 279$\.$55 \\
			\	\ 400 & 0$\.$24  & \ 0$\.$87 & \ 736$\.$24&&&&\ \ 400  & 0$\.$17  & \ 0$\.$74 & \ \ 577$\.$11\\
			\ 1000 & 0$\.$37 & \ 1$\.$58 & 1268$\.$74&&&&\ 1000 & 0$\.$37 & \ 1$\.$74 & \ 1439$\.$10\\
			\ 2000 & 0$\.$51 & \ 2$\.$98 & 2201$\.$27&&&&\ 2000 & 0$\.$79  & \ 3$\.$65 & \ 2975$\.$08\\	
			\ 3000 & 0$\.$85 & \ 5$\.$35  & 3121$\.$63&&&&\ 3000 & 1$\.$20  & \ 5$\.$61 & \ 4324$\.$52 \\		
			\ 5000 & 0$\.$81   & 11$\.$16 & 4806$\.$57&&&& \ 5000 & 2$\.$06  & \ 8$\.$42 & \ 7033$\.$50 \\	
			10000 & 1$\.$32 & 34$\.$41 & 9298$\.$11&&&&10000 & 5$\.$21 & 18$\.$53 & 15855$\.$51 \\
			\hline
		\end{tabular}}
	\end{table}

%Especially at the right hand of Table~\ref{Tab:T}, the computation time in case that $p$ increases with $n$ fixed are of the most interest in ultrahigh dimensional data analysis, survival impact index based screening still need some development to be applied in the realistic ultrahigh dimensional data analysis, considering its increasingly intolerable computation time with increasing of dimension $p$.
	%In all, with the increasing of the dimension $p$ or of the sample size $n$, the computation time which SII needs seems more intolerable. And SII need some development when applied in the realistic ultrahigh-dimensional data analysis. Meanwhile Table \ref{Tab:P} show that proposed \MDRSIS also performs better than SII in most cases when SII is well performed and efficient.

Next,  we consider $n=300$ and $p= 2000$. As SII is very time consuming for  $p=2000$ in simulation, we only compare \MDRSIS \ with QaSIS under this setting. To assess the effect of  $\alpha$ in QaSIS, we obtain results for QaSIS with $\alpha=0\.5$ and $0\.7$. The results reported in Table~\ref{Tab:P2000} show that  \MDRSIS \ overwhelms QaSIS regardless of the choices of $\alpha$. Moreover, the performance of QaSIS can be influenced by the choice of $\alpha$.
	\begin{table}
		\centering
		\caption{Simulation proportions of each  relevant covariate and all relevant covariates  selected by  \MDRSIS\ and QaSIS with $\alpha =0.5$ and $0.7$;  $p=2000$, $n=300$, $d_n = \lfloor n/ \log n\rfloor =52$, simulation replication  $500$}\label{Tab:P2000}
		\scalebox{0.78}{
			\begin{tabular}{ccccccccccccccccccc}%{|lr|rrrrr|rrrrrr|rrrrrr|c}
				\hline
				&& \multicolumn{5}{c}{\centering \MDRSIS}&& \multicolumn{5}{c}{\centering QaSIS ($\alpha=0\.5$)}&  &\multicolumn{5}{c}{\centering QaSIS ($\alpha=0\.7$)}\\
				\cline{3-7} \cline{9-13} \cline{15-19}%\hline			
				&& \multicolumn{4}{c}{\centering relevant covariate}&&&\multicolumn{4}{c}{\centering relevant covariate}&&&\multicolumn{4}{c}{\centering relevant covariate}&\\
				%&\quad& \multicolumn{4}{c}{\centering $\calP_s$}&$\calP_a$&\quad&\multicolumn{4}{c}{\centering $\calP_s$}&$\calP_a$& \quad&\multicolumn{4}{c}{\centering $\calP_s$}&$\calP_a$\\
				model&$\rho$&1&2&3&4& all&&1&2&3&4&all&&1&2&3&4&all\\
				%&$\x_{(1)}$&$\x_{(2)}$&$\x_{(3)}$&$\x_{(4)}$&All&\quad&$\x_{(1)}$&$\x_{(2)}$&$\x_{(3)}$&$\x_{(4)}$&All&\quad&$\x_{(1)}$&$\x_{(2)}$&$\x_{(3)}$&$\x_{(4)}$&All\\
				\hline
				M1& \ \ 0 & 0$\.$96 & 0$\.$96 & 0$\.$94 & 0$\.$93 & 0$\.$79 &   & 0$\.$26 & 0$\.$23 & 0$\.$78 & 0$\.$79 & 0$\.$04 &   & 0$\.$22 & 0$\.$22 & 0$\.$04 & 0$\.$06 & 0$\.$00 \\
				&0.4    & 1$\.$00 & 0$\.$99 & 0$\.$99 & 1$\.$00 & 0$\.$98 &   & 0$\.$37 & 0$\.$32 & 0$\.$84 & 0$\.$84 & 0$\.$09 &   & 0$\.$29 & 0$\.$29 & 0$\.$05 & 0$\.$06 &0.00 \\ 			
				&0.8    & 1$\.$00 & 1$\.$00 & 1$\.$00 & 1$\.$00 & 1$\.$00 &   & 0$\.$76 & 0$\.$78 & 0$\.$88 & 0$\.$86 & 0$\.$47 &   & 0$\.$76 & 0$\.$77 & 0$\.$24 & 0$\.$20 & 0$\.$06 \\
				M2&\ \ 0 & 0$\.$98 & 0$\.$99 & 0$\.$92 & 0$\.$93 & 0$\.$84 &   & 0$\.$39 & 0$\.$39 & 0$\.$49 & 0$\.$49 & 0$\.$05 &   & 0$\.$27 & 0$\.$26 & 0$\.$11 & 0$\.$16 & 0$\.$00 \\
				&0.4    & 1$\.$00 & 1$\.$00 & 1$\.$00 & 1$\.$00 & 1$\.$00 &   & 0$\.$42 & 0$\.$44 & 0$\.$75 & 0$\.$75 & 0$\.$12 &   & 0$\.$31 & 0$\.$31 & 0$\.$35 & 0$\.$32 & 0$\.$01 \\
				&0.8    & 1$\.$00 & 1$\.$00 & 1$\.$00 & 1$\.$00 & 1$\.$00 &   & 0$\.$80 & 0$\.$93 & 0$\.$96 & 0$\.$91 & 0$\.$66 &   & 0$\.$76 & 0$\.$89 & 0$\.$81 & 0$\.$72 & 0$\.$48 \\
				M3&\ \ 0  & 0$\.$89 & 0$\.$90 & 0$\.$92 & 0$\.$92 & 0$\.$66 &   & 0$\.$97 & 0$\.$98 & 0$\.$67 & 0$\.$67 & 0$\.$42 &   & 0$\.$09 & 0$\.$10 & 0$\.$25 & 0$\.$24 & 0$\.$00 \\
				&0.4    & 0$\.$95 & 0$\.$92 & 1$\.$00 & 1$\.$00 & 0$\.$88 &   & 0$\.$94 & 0$\.$92 & 0$\.$89 & 0$\.$91 & 0$\.$75 &   & 0$\.$05 & 0$\.$07 & 0$\.$41 & 0$\.$41 & 0$\.$00 \\
				&0.8    & 1$\.$00 & 1$\.$00 & 1$\.$00 & 1$\.$00 & 1$\.$00 &   & 0$\.$90 & 0$\.$93 & 0$\.$98 & 0$\.$97 & 0$\.$86 &   & 0$\.$11 & 0$\.$23 & 0$\.$74 & 0$\.$70 & 0$\.$03 \\
				M4&\ \ 0  & 0$\.$97 & 0$\.$98 & 0$\.$91 & 0$\.$84 & 0$\.$72 &   & 0$\.$60 & 0$\.$56 & 0$\.$82 & 0$\.$86 & 0$\.$26 &   & 0$\.$75 & 0$\.$74 & 0$\.$65 & 0$\.$33 & 0$\.$12 \\
				&0.4    & 0$\.$14 & 0$\.$92 & 1$\.$00 & 0$\.$97 & 0$\.$12 &   & 0$\.$21 & 0$\.$48 & 0$\.$97 & 0$\.$96 & 0$\.$10 &   & 0$\.$23 & 0$\.$64 & 0$\.$91 & 0$\.$44 & 0$\.$05 \\
				&0.8    & 0$\.$49 & 1$\.$00 & 1$\.$00 & 1$\.$00 & 0$\.$49 &   & 0$\.$62 & 0$\.$96 & 0$\.$99 & 0$\.$98 & 0$\.$61 &   & 0$\.$27 & 0$\.$94 & 0$\.$99 & 0$\.$92 & 0$\.$26 \\
				M5&\ \ 0  & 1.00 & 1.00 & 1.00 & 1.00 & 0.99 &   & 0.83 & 0.32 & 0.82 & 0.21 & 0.08 &   & 0.23 & 0.39 & 0.67 & 0.08 & 0.01 \\
				&0.4    & 1.00 & 1.00 & 1.00 & 1.00 & 1.00 &   & 0.85 & 0.54 & 0.86 & 0.27 & 0.14 &   & 0.40 & 0.68 & 0.80 & 0.10 & 0.03 \\
				&0.8    & 1.00 & 1.00 & 1.00 & 1.00 & 1.00 &   & 0.91 & 0.98 & 0.97 & 0.67 & 0.64 &   & 0.97 & 1.00 & 1.00 & 0.75 & 0.74 \\
				\hline
			\end{tabular}}
		\end{table}

Under model M5, only $C$ is related with $X_6$, which is denoted as relevant covariate 4 in Tables~\ref{Tab:P} and \ref{Tab:P2000}.
Neither SII nor QaSIS can capture $X_6$, whereas \MDRSIS \ selects $X_6$ with high probability. 
This is expected since SII and QaSIS are not developed to search covariates related with $C$.

		%%%%%%%%%%%%%%%%%%%%%%%%%%%%%%%%%%%%%%%%%%%%%%%%%%%%%%%%%%%%%%%%%%%%%%%%%%%%%%%%%%%%%%%%%%%%%%%%%%%%%%%%%%%%%%%

Now, we assess the performance of \MDRIS. From Tables~\ref{Tab:P} and \ref{Tab:P2000}, all the three methods under consideration performed not well in model M4 with $\rho=0\.4$ and $0\.8$, where the first relevant covariate is missed by \MDRSIS \ with high frequency. Since \MDRSIS \ performed well in model M4 with $\rho =0$, these results indicate that the phenomenon  described in  Proposition 4.1 occurs when $\rho \neq 0$.  
Thus, we run more simulations under M4 with $\rho = 0.4$ and 0.8 to show that \MDRIS\ picks up $X_1$ missed by \MDRSIS\ and hence improves the overall performance. To see the performance of \MDRIS\  when  \MDRSIS\ already has a satisfactory performance, we include model M3 with $\rho=0\.8$. 
Furthermore, we check the influence of $p_{1},\dots,p_{S}$, the sizes of  covariate sets in iteration steps, and $S$, the number of iterations. We include $S=2$ and $S=4$, nearly equal, increasing, and decreasing  $p_j$'s, with $\sum_j p_j = 
d_n  = \lfloor n/ \log n\rfloor$, which is 37 when $n=200$ and $52$ when $n=300$.   
The special case with  $p_1 = d_n $ and $p_2=p_3=p_4=0$ is  \MDRSIS\ without iteration.

\begin{table}[h]
	\centering
	\caption{Simulation proportions of all relevant covariates  selected by  \MDRSIS\ and \MDRIS \ with different sizes $p_j$'s; simulation replication  $500$
	}\label{Tab:ISvsSIS1}
	\scalebox{0.78}{
		\begin{tabular}{ccccccccc}
			\hline
			& & \multicolumn{4}{c}{sizes in iteration} & \multicolumn{3}{c}{model and $\rho$} \\
			& 			 &$p_{1}$&$p_{2}$&$p_{3}$&$p_{4}$& M4, $\rho = 0.4$&M4, $\rho=0.8$ &M3, $\rho=0.8$\\ 
			\hline
			$n=200$, $p=400$&		 & 37& 0 & 0 & 0 & 0.19 & 0.52 & 0.99 \\
			&	 & 26 & 11 & 0 & 0 & 0.69 & 0.80 & 0.97 \\ 
			& & 23 & 14 & 0 & 0 & 0.69 & 0.81 & 0.97 \\ 
			& & 19 & 18 & 0 & 0 & 0.65 & 0.84 & 0.96 \\ 
			& & 14 & 23 & 0 & 0 & 0.60 & 0.87 & 0.94 \\ 
			& & 11 & 26 & 0 & 0 & 0.53 & 0.86 & 0.88 \\ 
			& & 24 & 5 & 4 & 4 & 0.64 & 0.75 & 0.97 \\ 
			& & 17 & 7 & 7 & 6 & 0.59 & 0.78 & 0.95 \\ 
			& & 10 & 9 & 9 & 9 & 0.48 & 0.77 & 0.86 \\ 
			\hline
			$n=300$, $p=2000$ &	 &52 & 0 & 0 & 0 & 0.12 & 0.49 & 1.00 \\
			& 	 & 40 & 12 & 0 & 0 & 0.80 & 0.78 & 1.00 \\ 
			& & 32 & 20 & 0 & 0 & 0.80 & 0.83 & 1.00 \\ 
			& & 26 & 26 & 0 & 0 & 0.77 & 0.86 & 1.00 \\ 
			& & 20 & 32 & 0 & 0 & 0.74 & 0.87 & 0.99 \\ 
			& & 12 & 40 & 0 & 0 & 0.63 & 0.89 & 0.98 \\ 
			& & 31 & 7 & 7 & 7 & 0.76 & 0.77 & 1.00 \\ 
			& & 26 & 10 & 8 & 8 & 0.76 & 0.80 & 1.00 \\ 
			& & 13 & 13 & 13 & 13 & 0.64 & 0.80 & 0.99 \\ 
			\hline
	\end{tabular}}		
\end{table}

	%%%%%%%%%%%%%%%%%%%%%%%%%%%%%%%%%%%%%%%%%%%%%%%%%%%%%%%%%%%%%%%%%%%%%%%%%%%%%%%%%%%%%%%%%%%%%%%%%%%%%%%%%%%%%%%%%%%%%%%%%%%%%%%%%%%%%%%%%	

\begin{table}[h]
	\centering
	\caption{Simulation proportions of all relevant covariates  selected by  \MDRSIS, \MDRIS, and \MDRSS; sizes of screened covariate sets by \MDRSS; $S=2$, $p_1 \approx p_2$ for \MDRIS; 
		$B=100$, $n_s=\lfloor 4n/5\rfloor$, $\pi_0=0\.3$ for \MDRSS; 
		simulation replication  $500$
	}\label{Tab:SSfix}
	\scalebox{0.78}{
		\begin{tabular}{ccccccccccccc}
			\hline
			& &&\multicolumn{3}{c}{prob selecting all relevant covariates}&
			&\multicolumn{2}{c}{size of \MDRSS\ }&\\
			\cline{4-6}  \cline{8-9}
			&		model & $d_n$ &\MDRSIS&\MDRIS&\MDRSS&&MED &IQR\\
			\hline
			$n=200$, $p=400$ &		M4, $\rho =0.4$ \quad & 37 & 0$\.$19 &   0$\.$65  &     0$\.$67     & &26 & 3 \\
			&		M4, $\rho = 0.8$ \quad & 37 & 0$\.$52 &      0$\.$84  &      0$\.$89&   & 26  & 4\\
			&	M3, $\rho = 0.8$ \quad & 37 & 0$\.$99 &0$\.$96  &       0$\.$96 &   & 25  & 4 \\
			$n=300$, $p=2000$ &		M4, $\rho =0.4$	\quad 	& 52 & 0$\.$12 & 0$\.$77  &     0$\.$72 &   & 28 &  4 \\
			&		M4, $\rho = 0.8$ \quad & 52 & 0$\.$49 &     0$\.$86  &     0$\.$86 &    &29   & 4 \\
			&		M3, $\rho = 0.8$ \quad & 52 & 1$\.$00 &       1$\.$00  &    0$\.$99&    & 27   &3\\
			\hline
	\end{tabular}}	
	\begin{tablenotes}
		\item \footnotesize	MED: the median size of screened covariate set by \MDRSS
		\item  IQR: the inter-quartile range of size of screened covariate set by \MDRSS
	\end{tablenotes}
\end{table}			
	
	The simulation proportions that all relevant covariates are selected are reported in Table \ref{Tab:ISvsSIS1}. From Table~\ref{Tab:ISvsSIS1},  \MDRIS\ improves \MDRSIS\ when the latter does not performs well, and is slightly whose than  \MDRSIS\ when \MDRSIS \ already has a satisfactory  performance.  
Regarding the influence of different patterns of $p_{1},\dots,p_{S}$ and $S$ on \MDRIS, the results in Table~\ref{Tab:ISvsSIS1} show that $S=2$ with nearly equal $p_j$'s or a large $p_1$ have better performances and therefore are recommended. 
		
	Finally, we assess the performance of \MDRSS\. From Table~\ref{Tab:ISvsSIS1}, the proportions that all relevant covariates are selected by \MDRIS \ are in a satisfactory range. Thus, it is of  interest to see whether \MDRSS \ can reduce the size of screened covariate set without losing the power in selecting all relevant covariates. 
	Under the setting in Table~\ref{Tab:ISvsSIS1} with $S=2$ and nearly equal $p_1$ and $p_2$, in Table~\ref{Tab:SSfix} 
	we list the proportions of selecting all relevant covariates by \MDRSS \ with 
	$B=100$ subsamples of size $n_s=\lfloor 4n/5\rfloor$ without replacement and threshold value $\pi_0=0\.3$ as suggested by \citet{STABLESIS2011}.
	Similar results for $\pi_0 = 0.4$ are obtained but not shown here.
	The median size of screened covariate set by \MDRSS \ and 
	the inter-quartile range of sizes are also reported in Table~\ref{Tab:SSfix}.
	The screened covariate size of \MDRSIS\ and \MDRIS \ is $d_n$, which is fixed when $n$ is fixed and included in  Table~\ref{Tab:SSfix}.
	The results in  Table~\ref{Tab:SSfix} show that 
	\MDRSS\  maintains a satisfactory level of selecting all relevant covariates, and decreases the size of screened covariate set by 	 $29.7\%$ to $48.1\%$.

%\csubsection{A Real Data Application}
\subsection{A Real Data Application}		
We apply our proposed methods to the diffuse large-B-cell lymphoma microarray data in \cite{DLBCL data}. This data set consists of measurements on $p=7399$ genes from $240$ patients.  The censored survival time $T^o$  ranges from $0$ to $21\.8$ years.
Following \cite{DLBCLtrain2004}, we use data from  $n=160$ patients as the sample training data and data from the rest 80 patients as validation data.
			
\cite{DLBCLtrain2004} applied a supervised principal components (PC) method using the training data to select 17 genes from $p$ genes. Then they used validation data to  fit a Cox proportional hazards model in which the covariate effect is a linear combination of the 17 genes. Using training data, \cite{QaSIS2013} selected $ \lfloor n/ \log n \rfloor =\lfloor 160 / \log 160 \rfloor  = 31$ genes by applying  QaSIS with  $\alpha = 0.4$. 
%	 considering the nearly half censoring of the survival time data. 
Using validation data, they also fitted a Cox proportional hazards model with a linear combination of the $31$ selected genes as the covariate effect. 
	
Based on the same training data set, we selected  $\lfloor n/ \log n \rfloor = 31$ genes by applying SII and the proposed \MDRSIS, \MDRIS, and \MDRSS, and then fitted a Cox proportional hazards model  with a linear combination of the $31$ selected genes as the covariate effect, based on the validation data set. 
For \MDRIS\, we used $S=2$, $p_1=16$, and $p_2=15$. For \MDRSS\, we used $B=100$ subsamples with $n_s=n/2 = 80$ and $\pi_0=0.4$, which resulted in $30$ selected genes.
		
Table~\ref{Tab:DLBCL} shows $R^2$ statistics of Cox proportional hazards models and the associated p-values of  log-rank tests, calculated by using the models with covariates selected by  these six methods. 
The $R^2$ statistic for each model measures the percentage of variation in survival time that is explained by the model. Thus, when comparing models, one would prefer the model with a large $R^2$ statistic. 
It's clear that the three methods we proposed are better than the others in terms of $R^2$.
\begin{table}[h]
				\centering
				\caption{$R^2$ statistics and p-values for six methods based on the diffuse large-B-cell lymphoma microarray data}\label{Tab:DLBCL}
				\scalebox{0.8}{
				\begin{tabular}{lcc}
					\hline
					\multicolumn{1}{c}{method} &$R^2$& p-value\\
					\hline
					Supervised PC& 0$\.$113 &0$\.$001\\
					QaSIS($\alpha=0\.4$)& 0$\.$375 &0$\.$083\\
					SII& 0$\.$358 &0$\.$335\\
					\MDRSIS\ & 0$\.$506 & 0$\.$008\\
					\MDRIS\ & 0$\.$511 & 0$\.$046\\
					\MDRSS\ & 0$\.$502 & 0$\.$015\\
					\hline
				\end{tabular}}
\end{table}		
			
\begin{figure}	
	\begin{minipage}{1\linewidth}
		\begin{center}
		\includegraphics[scale=0.6]{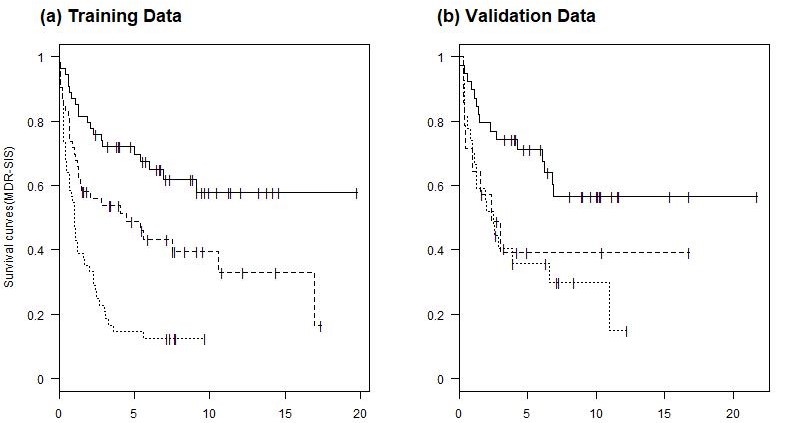}
		\end{center}
	\end{minipage}     
	\hfill
	\begin{minipage}{1\linewidth}
		\begin{center}
		\includegraphics[scale=0.6]{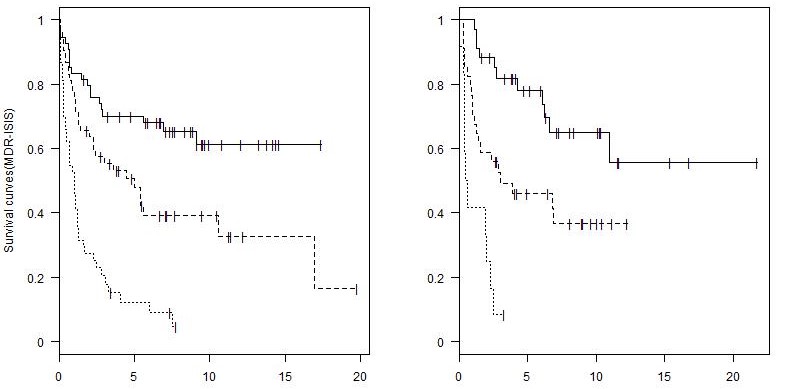}
	    \end{center}
	\end{minipage}
	\hfill
	\begin{minipage}{1\linewidth}
		\begin{center}
		\includegraphics[scale=0.6]{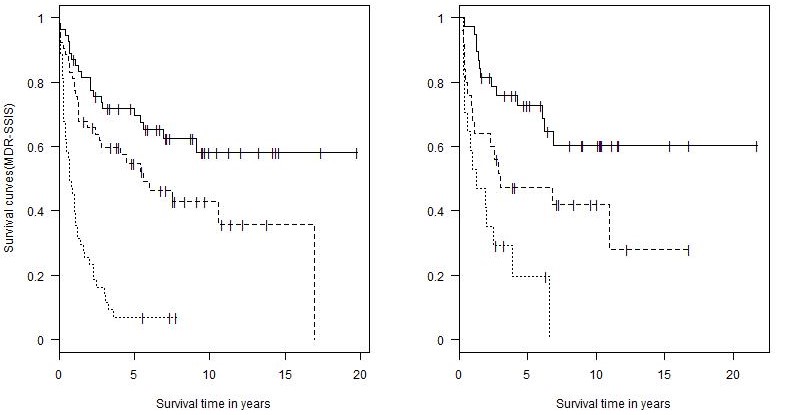}
	    \end{center}
		\caption{Kaplan-Meier estimates of survival curves for the low-risk patients group (solid), the intermediate-risk patients group (dash), and the high-risk patients group (small dash) based on \MDRSIS, \MDRIS, and \MDRSS; panel (a) is based on training data and panel (b) is based on  validation data}\label{pic:MDR}
	\end{minipage}
\end{figure}

We also evaluate the predictive performance of the proposed methods similarly with \cite{DLBCLtrain2004} and \cite{DLBCL2008}. A Cox proportional hazards model is fitted with these subsets of genes selected by the proposed method as the predictors. Three risk groups of patients, the low-risk patients, the intermediate-risk patients, and the high-risk patients, are defined according to the $33\%$ and $66\%$ quantiles of the estimated risk scores. Figure \ref{pic:MDR} is based on different subsets of genes selected by \MDRSIS, \MDRIS, and \MDRSS, respectively. Panel (a) in Figure \ref{pic:MDR} shows the Kaplan-Meier estimates of survival curves for the three risk groups of patients in the training data, whereas panel (b) shows the same curves based on  the validation data.

Panel (a) of Figure \ref{pic:MDR} shows that all three methods achieved good separation of the three risk groups, which indicates a good model fit to the training data. The log-rank test of difference among three survival curves yielded the p-value of 0 for all cases, which confirms our visual examination. The first block in panel (b) of Figure \ref{pic:MDR} shows that the estimator based on \MDRSIS\ separated the low-risk group with the intermediate and high-risk groups, resulting in a p-value of $0.0146$. However, it did not achieve satisfactory separation between the intermediate and high-risk groups. Meanwhile, the second and third blocks of  panel (b) of Figure \ref{pic:MDR} show that the estimators based on \MDRIS \ and \MDRSS\  achieved a better separation of the three risk groups with the validation data, resulting in a p-value 
nearly 0. 
Overall, our proposed methods in conjunction with a Cox proportional hazards model demonstrate competent variable screening and model fitting.

%\csection{Appendix: Proofs}\label{sec:proofs}
\section{Appendix: Proofs}\label{sec:proofs}

\noindent {\sc Proof of Proposition \ref{theo: svs and sdr}}. Let $\cala^c$ be the complement of $\cala$ in $\{1,...,p\}$ and $\I_{\cala^c}= \diag\{d_1,\dots, d_p\}$ be the $p\times p$ dimensional diagonal matrix with
$d_i = 1$ for $i\in \cala$ and $d_i = 0$ for $i\in \cala^c$. Similarly define $\I_{\cala}$ such that $\I_{\cala^c} + \I_{\cala} = \I_p$.
This proposition can be proved if we prove the equivalent result that
$i\in \cala^c$ if and only if $\e_i\trans\B=\0$.
First consider the “only if” part. By definition, $(\T, C) \indep \X_{\cala^c} \mid \X_{\cala}$. By the definition of $\B$, we have $\mathcal{S}_{(\T,\C)\mid \X}=\spn(\B)\subseteq\spn(\I_{\cala})$.
It follows immediately that $\I_{\cala^c}\B= 0$. For $i\in \cala^c$, the $i$th row of $\I_{\cala^c}$ is $\e_i\trans$. Thus we have $\e_i\trans\B=\0$. Now consider the “if” part. Take $\I_{\{i\}} = \diag\{\e_i\}$. Then $\e_i\trans\B= 0$ guarantees that $\I_{\{i\}}\B=0$. Let $\varepsilon = \{1,\dots, i-1, i+1,\dots , p\}$. Then $\B\trans\X=\B\trans\I_p\X=\B\trans\I_\varepsilon\X$. From the definition of $\mathcal{S}_{(\T,\C)\mid \X}$, we have $(\T, C) \indep \X \mid \B\trans\X$, which is $(\T, C) \indep \X \mid \B\trans\I_\varepsilon\X$. As $\I_\varepsilon\X$ involves only 0 and elements in $\X_\varepsilon$, we have $\mathcal{S}_{(\T,\C)\mid \X_\varepsilon}$. By the definition of the active set $\cala$, we know $\cala\subseteq\varepsilon$ and $i\in \cala^c$.
\eop
									
\noindent {\sc Proof of Proposition \ref{pro: Censored RD in CS}}. For part (i), denote $E[(\Z-\widetilde{\Z})(\Z-\widetilde{\Z})\trans\mid \T,\widetilde{\T},\C,\widetilde{\C}]$ by $\A(\T,\widetilde{\T},\C,\widetilde{\C})$ and $E[(\Z-\widetilde{\Z})(\Z-\widetilde{\Z})\trans\mid \T^o,\widetilde{\T^o},\delta,\widetilde{\delta}]$ by $\D(\T^o,\widetilde{\T^o},\delta,\widetilde{\delta})$. Let $\bfnu\in \mathcal{S}_{(\T,\C)\mid\Z}^{\perp}$ and denote the column space of a matrix $\W$ by $\spn(\W)$. $\spn[2\I_p-\D(\T^o,\widetilde{\T^o},\delta,\widetilde{\delta})]\subset \mathcal{S}_{(\T,\C)\mid\Z}$ leads to the fact that $\spn(\M) \subset\mathcal{S}_{(\T,\C)\mid\Z}$ cause $\bfnu\trans M\bfnu=E\{\bfnu\trans[2\I_p-\D(\T^o,\widetilde{\T^o},\delta,\widetilde{\delta})]^2\bfnu\}=0$. Thus, it suffices to prove $\spn[2\I_p-\D(\T^o,\widetilde{\T^o},\delta,\widetilde{\delta})]\subset \mathcal{S}_{(\T,\C)\mid\Z}$. 
									
First we prove that $\spn[2\I_p-\A(\T,\widetilde{\T},\C,\widetilde{\C})] \subset \mathcal{S}_{(\T,\C)\mid\Z}$ for any given $(\T,\widetilde{\T},\C,\widetilde{\C})$ under conditions (A1) and (A2). By choice of $(\widetilde{\Z},\widetilde{\T},\widetilde{\C})$, $(\Z,\T,\C)\indep (\widetilde{\Z},\widetilde{\T},\widetilde{\C})$. Thus
\begin{align}
\A(\T,\widetilde{\T},\C,\widetilde{\C})&=E[(\Z-\widetilde{\Z})(\Z-\widetilde{\Z})\trans\mid \T,\widetilde{\T},\C,\widetilde{\C}]\notag\\
&=E(\Z\Z\trans\mid \T,\C)-E(\Z\mid \T,\C)E(\widetilde{\Z}\trans\mid\widetilde{\T},\widetilde{\C})\notag\\
&~~~+E(\widetilde{\Z}\widetilde{\Z}\trans\mid\widetilde{\T},\widetilde{\C})-E(\widetilde{\Z}\mid\widetilde{\T},\widetilde{\C})E(\Z\trans\mid \T,\C).\label{eq:A expression}
\end{align}
It suffices to show that $\mathcal{S}_{(\T,\C)\mid\Z}^{\perp}\subset \{\spn[2\I_p-\A(\T,\widetilde{\T},\C,\widetilde{\C})]\}^{\perp}$.\\
By assumption (A1), $E(\bfnu\trans \Z\mid P\Z)=\ba\trans P\Z$ for some $\ba \in \R^p$. Because $\Z\Z\trans=\I$ and $\bfnu\bot P\Z\in \mathcal{S}_{(\T,\C)\mid\Z}^{\perp}$ we have
\begin{align*}
0=E(\bfnu\trans P\ba)=E\{E[\bfnu\trans \Z(\ba\trans P\Z)\trans\mid P\Z]\}=E[\ba\trans P\Z(\ba\trans P\Z)\trans]=E(\ba \trans P\ba).
\end{align*}
Thus $E^2(\bfnu\trans \Z\mid P\Z)=(\ba\trans P\Z)^2=\ba \trans P\ba=0$. By assumption (A2),
\begin{align*}
E[(\bfnu\trans \Z)^2\mid P\Z]=c+E^2(\bfnu\trans\Z\mid P\Z)=c,
\end{align*}
where $c$ is a constant. Take unconditional expectations on both sides to obtain $c=\bfnu\trans \bfnu$. Thus $E[(\bfnu\trans \Z)^2]=\bfnu\trans \bfnu$. Because $(\T, \C)\indep \Z\mid P\Z$, we have
\begin{align*}
E(\bfnu\trans \Z\mid \T,\C)&=E[E(\bfnu\trans \Z\mid P\Z)\mid \T,\C]=0,\\
E[(\bfnu\trans \Z)^2\mid \T,\C]&=E\{E[(\bfnu\trans \Z)^2\mid P\Z]\mid \T,\C\}=\bfnu\trans \bfnu.
\end{align*}
Substitute these in to (\ref{eq:A expression}), then the fact that $(\Z,\T,\C)$ and $(\widetilde{\Z},\widetilde{\T},\widetilde{\C})$ have the same distribution lead to $\bfnu\trans \A(\T,\widetilde{\T},\C,\widetilde{\C})\bfnu=2\bfnu\trans \bfnu$, implying that
\begin{align*}
\bfnu\trans[2\I_p-\A(\T,\widetilde{\T},\C,\widetilde{\C})]\bfnu=0.
\end{align*}
Thus $\spn[2\I_p-\A(\T,\widetilde{\T},\C,\widetilde{\C})]\subset\mathcal{S}_{\T,\C\mid\Z}$.
Finally by derivation of $\T^o$ and $\delta$, we have $E(\Z\mid\T^o,\delta)=E[E(\Z\mid \T,\C)\mid\T^o,\delta]$ and $E(\Z^2\mid\T^o,\delta)=E[E(\Z^2\mid \T,\C)\mid \T^o,\delta]$. Thus $(\widetilde{\Z},\widetilde{\T},\widetilde{\C})\indep (\Z,\T,\C)$ leads to $E[\A(\T,\widetilde{\T},\C,\widetilde{\C})\mid \T^o,\widetilde{\T^o},\delta,\widetilde{\delta}]=\D(\T^o,\widetilde{\T^o},\delta,\widetilde{\delta})$. Taking conditional expectation on $\A(\T,\widetilde{\T},\C,\widetilde{\C})$ given $(\T^o,\widetilde{\T^o},\delta,\widetilde{\delta})$, then we have $\spn[2\I_p-\D(\T^o,\widetilde{\T^o},\delta,\widetilde{\delta})]\subset \mathcal{S}_{(\T,\C)\mid\Z}$, which leads to the result $\spn(\M) \subset\mathcal{S}_{(\T,\C)\mid\Z}$.
									
For part (ii), with similar argument in proof of Theorem 3 in \cite{DR2007}, if $\spn(\M) \subset\mathcal{S}_{(\T,\C)\mid\Z}$, $\M=\M\trans$ and $\M\geq0$, then 
\begin{align}\label{M_CS}
\spn(\M) =\mathcal{S}_{(\T,\C)\mid\Z}
~~~\text{if and only if}~~~ \bfpsi\trans \M \bfpsi> 0 ~\text{for all}~ \bfpsi\in∈ \mathcal{S}_{(\T,\C)\mid\Z},~\bfpsi\neq 0.
\end{align}
Note that $\spn(\M) \subset\mathcal{S}_{(\T,\C)\mid\Z}$ is guaranteed by assumptions (A1) and (A2) , and $\M=\M\trans$ and $\M\geq0$ follow from the definition of $\M$.
									
Let $\bfpsi \in \mathcal{S}_{(\T,\C)\mid\Z}$ and $\bfpsi\neq0$. By (\ref{M_CS}), it suffices to show that $\bfpsi\trans \M \bfpsi> 0$. Without loss of generality, assume that $\parallel\bfpsi\parallel=1$. Write $\D(\T^o,\widetilde{\T^o},\delta,\widetilde{\delta}) - 2\I_p$ as
$\C(\T^o,\widetilde{\T^o},\delta,\widetilde{\delta})$. Then
\begin{align*}
\bfpsi\trans \M \bfpsi=\bfpsi\trans E[\C(\T^o,\widetilde{\T^o},\delta,\widetilde{\delta})(\I_p-\bfpsi\bfpsi\trans)\C(\T^o,\widetilde{\T^o},\delta,\widetilde{\delta})]\bfpsi+E[\bfpsi\trans \C(\T^o,\widetilde{\T^o},\delta,\widetilde{\delta})\bfpsi].
\end{align*}
Because $\I_p-\bfpsi\bfpsi\trans\geq 0$, the first term on the right is nonnegative. By assumption (A3), $\bfpsi\trans \D(\T^o,\widetilde{\T^o},\delta,\widetilde{\delta})\bfpsi$ is nondegenerate; thus $\bfpsi\trans \C(\T^o,\widetilde{\T^o},\delta,\widetilde{\delta})\bfpsi$ is nondegenerate. By Jensen's inequality, $E[(\bfpsi\trans \C(\T^o,\widetilde{\T^o},\delta,\widetilde{\delta})\bfpsi)^2]>[E(\bfpsi\trans \C(\T^o,\widetilde{\T^o},\delta,\widetilde{\delta})\bfpsi)]^2=0$, where the equality holds because $E\C(\T^o,\widetilde{\T^o},\delta,\widetilde{\delta}) = 0$.
\eop
								
\noindent {\sc Proof of Proposition \ref{prop:gkstar}}. Denote $\spn\{h(\D_{ijlm}):ijlm\}$ by $\spn\{h(\D_{ijlm}): i, l=0~\text{or}~1,~j=1,\dots, H_i,~m=1,\dots, H_l\}$. Note that  $g_k^*=\sum_{ij}\sum_{lm}p_{ij}p_{lm}[\e_k\trans{\bSig}^{-1/2}$ $(2\I_p-\D_{ijlm}){\bSig}^{-1/2}\e_k]^2$ and $\G=\sum_{ij}\sum_{lm}p_{ij}p_{lm}(2\I_p-\D_{ijlm})^2$.
By Proposition \ref{pro: Censored RD in CS}, (A1), (A2) and (A3) guarantee $\spn(\G)=\calS_{(\T,\C) \mid\Z}$. By the invariance law of the central space, we have $\spn({\bSig}^{-1/2}\G{\bSig}^{-1/2})=\calS_{(\T,\C) \mid\X}$.
If $k\in \cala^c$, we know from Lemma A.2 in \cite{TestDR2016} that $\e_k\trans{\bSig}^{-1/2}\G{\bSig}^{-1/2}\e_k=0$. Because $p_{i,j}p_{l,m}>0$, $\e_k\trans{\bSig}^{-1/2}(2\I_p-\D_{ijlm})=\0$ for any $(i,j)$ and $(l,m)$. From the expression of $g_k^*$, we have $g_k^*=0$ if $k\in \cala^c$.
Condition (A3) guarantees that $\spn\{(2\I_p-\D_{ijlm})^2:ijlm\}=\calS_{(\T,\C) \mid\Z}$, which in turn implies $\spn\{2\I_p-\D_{ijlm}:ijlm\}=\calS_{(\T,\C) \mid\Z}$. From the invariance law of the central space, $\e_k\trans{\bSig}^{-1/2}(2\I_p-\D_{ijlm}){\bSig}^{-1/2}\e_k>0$ for at least one set of $(i,j)$ and $(l,m)$ if $k\in \cala$. Otherwise we get a
contradiction to the “only if” part of Lemma A.2 in \cite{TestDR2016}. Thus, we have
$g_k^*\geq p_{ij}p_{lm}[\e_k\trans{\bSig}^{-1/2}(2\I_p-\D_{ijlm}){\bSig}^{-1/2}\e_k]^2>0$ if $k\in \cala$.
\eop
									
\noindent {\sc Proof of Lemma \ref{Lemma: rewriten gkstar}}.
Note that   $\sum_{ij}\sum_{lm}p_{ij}p_{lm}[\e_k\trans{\bSig}^{-1/2}(2\I_p-\D_{ijlm}){\bSig}^{-1/2}\e_k]^2$ is the discretized version of $E[\{\e_k\trans{\bSig}^{-1/2}(2\I_p-\D(\T^o,\widetilde{\T^o},\delta,\widetilde{\delta})){\bSig}^{-1/2}\e_k\}^2]$, $2\sum_{lj} p_{lj}[\e_k\trans{\bSig}^{-1/2}(p_{lj}^{-1}$ $\V_{lj}-\I_p){\bSig}^{-1/2}\e_k]^2$ is the discretized version of $2E[\{\e_k\trans{\bSig}^{-1/2}[E(\Z\Z\trans\mid\T^o,\delta)-\I_p]{\bSig}^{-1/2}\e_k\}^2]$ and $4\left(\sum_{lj} p_{lj}^{-1}\e_k\trans{\bSig}^{-1/2}\U_{lj}\trans{\bSig}^{-1/2}\e_k\right)^2$ is the discretized version of $4(E[\e_k\trans{\bSig}^{-1/2}E(\Z\mid\T^o,\delta)$ $E\trans(\Z\mid\T^o,\delta){\bSig}^{-1/2}\e_k])^2$.
Let $a_k(\T^o,\widetilde{\T^o},\delta,\widetilde{\delta})= \e_k\trans{\bSig}^{-1/2}$ $(2\I_p-\D(\T^o,\widetilde{\T^o},\delta,\widetilde{\delta})){\bSig}^{-1/2}\e_k$, all we need to prove
is that
\begin{align}\label{eq:Eak2}
E[a_k^2(\T^o,\widetilde{\T^o},\delta,\widetilde{\delta})]&=2E[\{\e_k\trans{\bSig}^{-1/2}[E(\Z\Z\trans\mid\T^o,\delta)-\I_p]{\bSig}^{-1/2}\e_k\}^2]\notag\\
&~~~~+4\left(E[\e_k\trans{\bSig}^{-1/2}E(\Z\mid\T^o,\delta)E(\Z\mid\T^o,\delta)\trans{\bSig}^{-1/2}\e_k]\right)^2\notag\\
&=2E[\{\e_k\trans{\bSig}^{-1/2}E(\Z\Z\trans\mid\T^o,\delta){\bSig}^{-1/2}\e_k\}^2]-2(\e_k\trans{\bSig}^{-1}\e_k)^2\notag\\
&~~~~+4\left(E[\e_k\trans{\bSig}^{-1/2}E(\Z\mid\T^o,\delta)E(\Z\mid\T^o,\delta)\trans{\bSig}^{-1/2}\e_k]\right)^2.
\end{align}
Let $d_k(\T^o,\widetilde{\T^o},\delta,\widetilde{\delta})=\e_k\trans{\bSig}^{-1/2}\D(\T^o,\widetilde{\T^o},\delta,\widetilde{\delta}){\bSig}^{-1/2}\e_k$. Then
\begin{align}\label{eq:Eak2dk2}
E[a_k^2(\T^o,\widetilde{\T^o},\delta,\widetilde{\delta})]=E[d_k^2(\T^o,\widetilde{\T^o},\delta,\widetilde{\delta})]-4(\e_k\trans{\bSig}^{-1}\e_k)^2.
\end{align}
With similar argument of Proposition \ref{pro: Censored RD in CS} (i), we have $d_k^2(\T^o,\widetilde{\T^o},\delta,\widetilde{\delta})=c_k(\T^o,\widetilde{\T^o},\delta,\widetilde{\delta})+c_k(\widetilde{\T^o},\T^o,\widetilde{\delta},\delta)$, where
\begin{align*}
c_k(\T^o,\widetilde{\T^o},\delta,\widetilde{\delta})=\e_k\trans{\bSig}^{-1/2}E(\Z\Z\trans\mid\T^o,\delta){\bSig}^{-1/2}\e_k-\e_k\trans{\bSig}^{-1/2}E(\Z\mid\T^o,\delta)E\trans(\widetilde{\Z}\mid\widetilde{\T^o},\widetilde{\delta}){\bSig}^{-1/2}\e_k;\\
c_k(\widetilde{\T^o},\T^o,\widetilde{\delta},\delta)=\e_k\trans{\bSig}^{-1/2}E(\widetilde{\Z}\widetilde{\Z}\trans\mid\widetilde{\T^o},\widetilde{\delta}){\bSig}^{-1/2}\e_k-\e_k\trans{\bSig}^{-1/2}E(\widetilde{\Z}\mid\widetilde{\T^o},\widetilde{\delta})E\trans(\Z\mid\T^o,\delta){\bSig}^{-1/2}\e_k.
\end{align*}
Plug them into (\ref{eq:Eak2}), it follows that
\begin{align}\label{eq:akck}
E[a_k^2(\T^o,\widetilde{\T^o},\delta,\widetilde{\delta})]=2E[c_k^2(\T^o,\widetilde{\T^o},\delta,\widetilde{\delta})]+2E[c_k(\T^o,\widetilde{\T^o},\delta,\widetilde{\delta})c_k(\widetilde{\T^o},\T^o,\widetilde{\delta},\delta)]-4(\e_k\trans{\bSig}^{-1}\e_k)^2.
\end{align}
By calculation, we have $E[c_k^2(\T^o,\widetilde{\T^o},\delta,\widetilde{\delta})]=c_{1k}+c_{2k}-c_{3k}-c_{4k}$, where
\begin{align*}
c_{1k}&=E[\{\e_k\trans{\bSig}^{-1/2}E(\Z\Z\trans\mid\T^o,\delta){\bSig}^{-1/2}\e_k\}^2];\\
c_{2k}&=E[\e_k\trans{\bSig}^{-1/2}E(\Z\mid\T^o,\delta)E\trans(\widetilde{\Z}\mid\widetilde{\T^o},\widetilde{\delta}){\bSig}^{-1/2}\e_k\e_k\trans{\bSig}^{-1/2}E(\Z\mid\T^o,\delta)E\trans(\widetilde{\Z}\mid\widetilde{\T^o},\widetilde{\delta}){\bSig}^{-1/2}\e_k];\\
c_{3k}&=E[\e_k\trans{\bSig}^{-1/2}E(\Z\mid\T^o,\delta)E\trans(\widetilde{\Z}\mid\widetilde{\T^o},\widetilde{\delta}){\bSig}^{-1/2}\e_k\e_k\trans{\bSig}^{-1/2}E(\Z\Z\trans\mid\T^o,\delta){\bSig}^{-1/2}\e_k];\\
c_{4k}&=E[\e_k\trans{\bSig}^{-1/2}E(\Z\Z\trans\mid\T^o,\delta){\bSig}^{-1/2}\e_k\e_k\trans{\bSig}^{-1/2}E(\Z\mid\T^o,\delta)E\trans(\widetilde{\Z}\mid\widetilde{\T^o},\widetilde{\delta}){\bSig}^{-1/2}\e_k].
\end{align*}
Because $(\Z,\T^o,\delta)\indep(\widetilde \Z,\widetilde{\T^o},\widetilde{\delta})$ and $E(\Z)=0$, we have $c_{3k}=c_{4k}=0$, $c_{2k}=E^2[\e_k\trans{\bSig}^{-1/2}E(\Z\mid\T^o,\delta)E\trans(\Z\mid\T^o,\delta){\bSig}^{-1/2}\e_k]$ and $E[c_k(\T^o,\widetilde{\T^o},\delta,\widetilde{\delta})c_k(\widetilde{\T^o},\T^o,\widetilde{\delta},\delta)]=(\e_k\trans{\bSig}^{-1}\e_k)^2+c_{2k}$.
Plug them into (\ref{eq:akck}), it follows to (\ref{eq:Eak2}) that complete the proof.
\eop
									
\noindent {\sc Proof of Proposition \ref{prop1_m1}}.
Let $\bfLambda=\sum_{ij}\sum_{lm}p_{ij}p_{lm}{\bSig}^{1/2}(2\I_p-\D_{ijlm}){\bSig}^{1/2}$. By propsition \ref{pro: Censored RD in CS}, (A1), (A2) and (A3) guarantee $\spn({\bSig}^{-1/2}\bfLambda{\bSig}^{-1/2})=\calS_{(\T,\C)\mid\Z}$. By the invariance law of the central space, we have $\spn({\bSig}^{-1}\bfLambda{\bSig}^{-1})=\calS_{(\T,\C) \mid\X}$. Let $\{\bfbeta_1,\dots,$ $\bfbeta_d\}=\B$ be a basis for $\cals_{(\T,\C)\mid\X}$. Then $\spn({\bSig}^{1/2}(2\I_p-\D_{ijlm}){\bSig}^{1/2})=\spn\{\bfzeta_1,\dots,\bfzeta_d\}$, where $\bfzeta_i={\bSig}\bfbeta_i$ for $i=1,\dots,d$. And $\sigma_k^4g_k=\sum_{ij}\sum_{lm}p_{ij}p_{lm}[\e_k\trans{\bSig}^{1/2}$ $(2\I_p-\D_{ijlm}){\bSig}^{1/2}\e_k]^2$. Similar with the proof of Proposition \ref{prop:gkstar}, if $\e_k\trans\bfzeta\neq0$ when $k\in \cala$, then $\e_k\trans{\bSig}^{1/2}(2\I_p-\D_{ijlm}){\bSig}^{1/2}\e_k>0$ for at least one set of $(i,j)$ and $(l,m)$. Thus, we have
$g_k\geq \sigma_k^{-4}p_{ij}p_{lm}[\e_k$ $\trans{\bSig}^{-1/2}(2\I_p-\D_{ijlm}){\bSig}^{-1/2}\e_k]^2>0$. So all we need to show is $\zeta_{ik}\neq 0$ for at least one of $i = 1,\dots,d$. Note that $\bfbeta_h \in \calS_{(\T,\C) \mid\X}$. By the definition of the active set $\cala$ and the central space $\calS_{(\T,\C) \mid\X}$, we have $\beta_{hj} = 0$ for $j \in \cala^c$. Thus, the $k$th component of $\bfzeta_h={\bSig}\bfbeta_h$ becomes $\zeta_{hk}=\sum_{j\in \cala}Cov(x_k,x_j)\beta_{hj}$.
If $k\in\cala$, then $Cov(x_k,x_j)\beta_{hj}$ has the same sign for all $j \in \cala$. Thus we have $\zeta_{hj}\neq0$ and $g_k > 0$ as a result.
\eop
									
\noindent {\sc Proof of Theorem \ref{theo: sure screening}}.
For part (i), let $C_1=2+\tau+\varsigma^{-1}K_0^2$ and $C_2=2+\tau+\varsigma^{-1}e^2K_0^2$. %From the proof of Theorems 1(a) and 4(a) in Cai, Liu and Luo (2011), we know that
%\begin{align}\label{eq:bound1}
%\p\{\max_{1\le k\le p}|\widehat\sigma_k^2-\sigma_k^2|\ge &\varsigma^{-1}C_1(\log p /n )^{1/2}\} \le 2 p^{-\tau-1},\\\label{eq:bound2}
%\p\{\max_{1\le k \le p}|\widehat\mu_k-\mu_k|\ge &\varsigma^{-1}C_2(\log p /n )^{1/2}\} \le 2 p^{-\tau-1}.
%\end{align}
By condition (C2), we see that $E\{\exp[t(z_k)^2I(\delta =l,\T^o\in I_{lj})]\}\le E\{\exp[t(z_k)^2]\} \le K_0$ and $E\{\exp[t|z_kI(\delta =l,\T^o\in I_{lj})|]\}\le E\{\exp(|tz_k|)\} \le e K_0$ for $|t|\le \varsigma$. Following similar arguments in the proof of Theorems 1(a) and 4(a) in \cite{CAI2011}, we derive that
\begin{align}
\p\{|\widehat V_{ljk}-V_{ljk}|\ge &\varsigma^{-1}C_1(\log p /n )^{1/2}\} \le 2 p^{-\tau-2},\label{eq:bound V}\\
\p\{|\widehat U_{ljk}-U_{ljk}|\ge &\varsigma^{-1}C_2(\log p /n )^{1/2}\} \le 2 p^{-\tau-2}.\label{eq:bound U}
\end{align}
Let $p_{\min}=\min\{p_{11},\ldots,p_{1H_1},p_{01},\ldots,p_{0H_0}\}$.  Note that $|I(\delta_i =l,y_i\in I_{lj})-p_{lj}|<1$, $E[I(\delta_i =l,y_i\in I_{lj})-p_{lj}]=0$ and $E[I(\delta_i =l,y_i\in I_{lj})-p_{lj}]^2=(1-p_{lj})p_{lj}\leq1/4$. By the Bernstein inequality (Lemma 2.2.9, \cite{EMPIRICALPROCESSES:1996}), we have
\begin{align}
&\p\left\{\left|\widehat p_{lj}-p_{lj}\right|\ge (2+\tau)(\log p/n)^{1/2} \right\}
=\p\{|\sum_{i=1}^nI(\delta_i =l,y_i\in I_{lj})-p_{lj}|\notag\\
&\ge (2+\tau)(n\log p)^{1/2} \}\le 2\exp\left\{-\frac{(2+\tau)^2n\log p}{2[n/4+(2+\tau)(n\log p)^{1/2}/3]}\right\}\le 2 p^{-\tau-2}.\notag
\end{align}
By condition (C1), we can assume that $\log p/n\le p_{\min}^2/(4+2\tau)^2<1/4$. Then 
\begin{align}
&\p\{|\widehat p^{-1}_{lj}-p^{-1}_{lj}|\ge (4+2\tau)p_{\min}^{-2}(\log p/n)^{1/2} \}\notag\\
&\le\p\{|\widehat p_{lj}-p_{lj}|\ge (2+\tau)(n\log p)^{1/2}\}+ \p\{p_{lj}\widehat p_{lj}\le p_{\min}(p_{\min}-p_{\min}/2) \}\notag\\
&\le 2p^{-\tau-2}+2p^{-\tau-2}=4p^{-\tau-2}.\label{eq:bound p}
\end{align}
By condition (C2), $E(z_k^2)=E(\varsigma z_k^2)/\varsigma \leq \varsigma^{-1}E[\exp(\varsigma z_k^2)]\leq \varsigma^{-1}K_0$. By Jensen's inequality, $E|z_k|\leq [E(z_k^2)]^{1/2}\leq \varsigma^{-1/2}K_0^{1/2}$. Thus $\max_{1\le k\le p}|U_{ljk}|\le\max_{1\le k\le p} E|z_k| \le \varsigma^{-1/2}K_0^{1/2}$ and  $\max_{1\le k\le p} V_{ljk}\le\max_{1\le k\le p} E(z_k^2)\le \varsigma^{-1}K_0$.
Let $C_3=(4+2\tau)\varsigma^{-1}K_0 p_{\min}^{-2}+4p^{-1}_{\min} K_0^{1/2}\varsigma^{-3/2}C_2+3(4+2\tau)^{-1}\varsigma^{-2}C_2^2.$ By $\log p/n\le p_{\min}^2/(4+2\tau)^2$. Then
\pagebreak
\begin{align*}
C_3&=(4+2\tau)\varsigma^{-1}K_0p_{\min}^{-2}+[2p^{-1}_{\min}+2(4+2\tau)p_{\min}^{-2}p_{\min}/(4+2\tau)]K_0^{1/2}\varsigma^{-3/2}C_2\\
&\quad+\{p_{\min}^{-1}p_{\min}/(4+2\tau)+(4+2\tau)p_{\min}^{-2}[p_{\min}/(4+2\tau)]^2\} \varsigma^{-2}C_2^2\\
&\geq (4+2\tau)\varsigma^{-1}K_0p_{\min}^{-2}+[2p^{-1}_{\min}+2(4+2\tau)p_{\min}^{-2}(\log p/n)^{1/2}]K_0^{1/2}\varsigma^{-3/2}C_2\\
&\quad+[p_{\min}^{-1}(\log p/n)^{1/2}+(4+2\tau)p_{\min}^{-2}\log p/n] \varsigma^{-2}C_2^2.
\end{align*}
Combining (\ref{eq:bound U}) and (\ref{eq:bound p}) together, we have
\begin{align}\nonumber
&\p\{|\widehat U_{ljk}^2/\widehat p_{lj}-U^2_{ljk}/p_{lj}|\ge C_3(\log p/n)^{1/2}\} \notag\\
&\le  \p\{|(\widehat p^{-1}_{lj}-p^{-1}_{lj}) U_{ljk}^2|\ge (4+2\tau)\varsigma^{-1}K_0 p_{\min}^{-2} (\log p/n)^{1/2}\}\notag\\
&~~~+\p\{|p_{lj}^{-1}(\widehat U_{ljk}-U_{ljk})^2|\ge p_{\min}^{-1} \varsigma^{-2}C_2^2\log p /n \}\notag\\
&~~~+\p\{|2p^{-1}_{lj} U_{ljk}(\widehat U_{ljk}-U_{ljk})|\ge 2\varsigma^{-3/2}K_0^{1/2}p^{-1}_{\min}C_2(\log p /n )^{1/2}\}\notag\\
%\le& \p\{|2(\widehat p^{-1}_{\ell}-p^{-1}_{\ell})(\widehat U_{\ell,k}-U_{\ell,k}) U_{\ell,k}|\ge 4K_0p_{\min}^{-2} n^{-1/2}\varsigma^{-1}C_2(\log p /n )^{1/2}\}\\
&~~~+ \p\{|2(\widehat p^{-1}_{lj}-p^{-1}_{lj})(\widehat U_{ljk}-U_{ljk}) U_{ljk}|\ge 2\varsigma^{-3/2}K_0^{1/2}(4+2\tau)p_{\min}^{-2} C_2\log p /n \}\notag
\end{align}
\begin{align}\nonumber
&~~~+ \p\{|(\widehat p^{-1}_{lj}-p^{-1}_{lj})(\widehat U_{ljk}-U_{ljk})^2|\ge (4+2\tau)p_{\min}^{-2} \varsigma^{-2}C_2^2(\log p /n )^{3/2}\}\notag\\
&\le  4p^{-\tau-2}+ 2p^{-\tau-2}+2p^{-\tau-2}+(4p^{-\tau-2}+2p^{-\tau-2})+(4p^{-\tau-2}+2p^{-\tau-2})\notag\\
&=20p^{-\tau-2}.\label{eq:bound sir}
\end{align}
Defines two positive constants  $C_4=(4+2\tau)\varsigma^{-2}K_0^2 p_{\min}^{-2}+4p^{-1}_{\min}K_0\varsigma^{-2}C_1+2(4+2\tau)^{-1}\varsigma^{-2}C_1^2$ and $C_5=(4+2\tau)\varsigma^{-1}K_0 p_{\min}^{-2}+[(4+2\tau)^{-1}+p_{\min}^{-1}] \varsigma^{-1}C_1$. By $\log p/n\leq [p_{\min}/(4+2\tau)]^2$, 
\begin{align*}
C_4&=(4+2\tau)\varsigma^{-2}K_0^2 p_{\min}^{-2}+[2p^{-1}_{\min}+(8+4\tau)p_{\min}^{-2}p_{\min}/(4+2\tau)] K_0\varsigma^{-2}C_1\\
&\quad+\{p_{\min}^{-1}p_{\min}/(4+2\tau)+(4+2\tau)p_{\min}^{-2}[p_{\min}/(4+2\tau)]^2\} \varsigma^{-2}C_1^2\\
&\geq (4+2\tau)\varsigma^{-2}K_0^2 p_{\min}^{-2}+[2p^{-1}_{\min}+(8+4\tau)p_{\min}^{-2}(\log p/n)^{1/2}] K_0\varsigma^{-2}C_1\\
&\quad+[p_{\min}^{-1}(\log p/n)^{1/2}+(4+2\tau)p_{\min}^{-2}\log p/n] \varsigma^{-2}C_1^2
\end{align*}
and
\begin{align*}
C_5&=(4+2\tau)\varsigma^{-1}K_0 p_{\min}^{-2}+p_{\min}^{-1} \varsigma^{-1}C_1+(4+2\tau)p_{\min}^{-2}(p_{\min}/2)^2 \varsigma^{-1}C_1\\
&\geq (4+2\tau)\varsigma^{-1}K_0 p_{\min}^{-2}+p_{\min}^{-1} \varsigma^{-1}C_1+(4+2\tau)p_{\min}^{-2} \varsigma^{-1}C_1 \log p/n.
\end{align*}
Similar to the derivation of (\ref{eq:bound sir}), by combining (\ref{eq:bound V}) and (\ref{eq:bound p}) we obtain 
\begin{align}
&\p\{|\widehat V_{ljk}^2/\widehat p_{lj}-V^2_{ljk}/p_{l,j}|\ge C_4(\log p/n)^{1/2}\} \notag\\
&\le  \p\{|(\widehat p^{-1}_{l,j}-p^{-1}_{lj}) V_{ljk}^2|\ge (4+2\tau)\varsigma^{-2}K_0^2 p_{\min}^{-2} (\log p/n)^{1/2}\}\notag\\
&~~~+\p\{|p_{lj}^{-1}(\widehat V_{ljk}-V_{ljk})^2|\ge p_{\min}^{-1} \varsigma^{-2}C_1^2\log p /n \}\notag\\
&~~~+\p\{|2p^{-1}_{lj} V_{ljk}(\widehat V_{ljk}-V_{ljk})|\ge 2\varsigma^{-2}K_0p^{-1}_{\min}C_1(\log p /n )^{1/2}\}\notag\\
%\le& \p\{|2(\widehat p^{-1}_{\ell}-p^{-1}_{\ell})(\widehat U_{\ell,k}-U_{\ell,k}) U_{\ell,k}|\ge 4K_0p_{\min}^{-2} n^{-1/2}\varsigma^{-1}C_2(\log p /n )^{1/2}\}\\\nonumber
&~~~+ \p\{|2(\widehat p^{-1}_{lj}-p^{-1}_{lj})(\widehat V_{ljk}-V_{ljk}) V_{ljk}|\ge (8+4\tau)\varsigma^{-2}K_0p_{\min}^{-2} C_1\log p /n \}\notag\\
&~~~+ \p\{|(\widehat p^{-1}_{lj}-p^{-1}_{lj})(\widehat V_{ljk}-V_{ljk})^2|\ge (4+2\tau)p_{\min}^{-2} \varsigma^{-2}C_1^2(\log p /n )^{3/2}\}\notag
\end{align}
\begin{align}
&\le  4p^{-\tau-2}+ 2p^{-\tau-2}+2p^{-\tau-2}+(4p^{-\tau-2}+2p^{-\tau-2})+(4p^{-\tau-2}+2p^{-\tau-2})\notag\\
&=20p^{-\tau-2}\label{eq:bound dr1}
\end{align}
and
\begin{align}
&\p\{|\widehat V_{ljk}/\widehat p_{lj}-V_{ljk}/p_{lj}|\ge C_{5}(\log p/n)^{1/2}\} \notag\\
&\le\p\{|(\widehat p^{-1}_{lj}-p^{-1}_{lj}) V_{ljk}|\ge (4+2\tau)\varsigma^{-1}K_0 p_{\min}^{-2}(\log p /n )^{1/2}\}\notag\\
&~~~+\p\{|p_{lj}^{-1}(\widehat V_{ljk}-V_{ljk})|\ge p_{\min}^{-1} \varsigma^{-1}C_1(\log p /n )^{1/2}\}\notag\\
&~~~+\p\{|(\widehat p^{-1}_{lj}-p^{-1}_{lj})(\widehat V_{ljk}-V_{ljk})^2|\ge (4+2\tau)p_{\min}^{-2} \varsigma^{-1}C_1(\log p /n )^{3/2}\}\notag\\
&\le 4p^{-\tau-2}+ 2p^{-\tau-2}+(4p^{-\tau-2}+2p^{-\tau-2})=12p^{-\tau-2}\label{eq:bound dr2}.
\end{align}
Let $H=H_0+H_1$. Define positive constant $C_0$ as follows:
\pagebreak
\begin{align}\label{def: C0}
C_0=H[2C_4+4C_5+8C_3+4C_3^2p_{\min}/(4+2\tau)].
\end{align}
Note that $\sum_{lj} U_{ljk}^2/p_{lj}=\cov[E(z_k\mid\delta,\T^o)]\le \var (z_k)=1$. By (\ref{eq:bound sir}), (\ref{eq:bound dr1}),  (\ref{eq:bound dr2}) and (\ref{def: C0}), we could derive that
\begin{align}
&\p\{|\widehat g_k-g_k|\ge C_0(\log p/n)^{1/2}\}\notag\\
&\le \p\{ |2\widehat V_{ljk}^2/\widehat p_{lj}-2V^2_{ljk}/p_{lj}|\ge 2C_4(\log p/n)^{1/2}\}\notag\\
&~~~+\p\{ |4\widehat V_{ljk}/\widehat p_{lj}-4V_{ljk}/p_{lj}|\ge 4C_{5}(\log p/n)^{1/2}\}\notag\\
&~~~+\p\{ |8(\sum_{lj} \widehat U_{ljk}^2/\widehat p_{lj}-\sum_{lj} U_{ljk}^2/p_{lj})(\sum_{lj} U_{ljk}^2/p_{lj})|\ge 8C_3(\log p/n)^{1/2}\}\notag\\
&~~~+\p\{|4( \sum_{lj} \widehat U_{ljk}^2/\widehat p_{lj}-\sum_{lj} U_{ljk}^2/p_{lj})^2|\ge 4C_3^2\log p/n\}\notag\\
&\le  20p^{-\tau-2}+12p^{-\tau-2}+20p^{-\tau-2}+20p^{-\tau-2}= 72p^{-\tau-2}.\label{eq:Theorem a}
\end{align}
Thus
\begin{align*}
\p\{\max_{1\le k\le p}|\widehat g_k-g_k|\ge C_0(\log p/n)^{1/2}\}\le p\max_{1\le k\le p}\p\{|\widehat g_k-g_k|\ge C_0(\log p/n)^{1/2}\}= 72p^{-\tau-1}.
\end{align*}
The proof of Theorem \ref{theo: sure screening} (i) is completed.
For part (ii), if $\cala\nsubseteq\widehat\cala$, then there must exist some $k\in \cala$ such that $\widehat g_k < c_0n^{-\kappa}$. It follows from condition (C3) that $|\widehat g_k-g_k|>c_0n^{-\kappa}$ for some $k\in \cala$. Let $p_0$ denotes the size of $\cala$. Thus
\begin{align}
\p(\cala\subseteq\widehat\cala )&\ge 1-\p\{|\widehat g_k-g_k|> c_0n^{-\kappa} \text{ for some } k\in \cala\}\notag\\
&\ge 1-\sum_{k=1}^{p_0}\p\{|\widehat g_k-g_k|> c_0n^{-\kappa}\}\notag\\
&\ge 1-p_0\max_{1\leq k\leq p}\p\{|\widehat g_k-g_k|> c_0n^{-\kappa}\}\notag\\
&\ge 1-p_0\max_{1\leq k\leq p}\p\{|\widehat g_k-g_k|> C_0(\log p/n)^{1/2}\},\label{eq:Theorem b}
\end{align}
where the last inequality follows from condition (C1). By (\ref{eq:Theorem a}), (\ref{eq:Theorem b}) and condition (C1), it follows that $\p(\cala\subseteq\widehat\cala )\ge 1-p_072p^{-\tau-2}$. By the definition of $\cala$ in (\ref{model csvs}), we have $p_0<p$ that lead to the final result $\p(\cala\subseteq\widehat\cala )\ge 1-72p^{-\tau-1}$.
\eop

%\noindent {\sc Proof of Proposition \ref{prop:A=B}}.
%Because $\calb$ is the smallest covariate set corresponding to all relevant predictors for the response $\T$. So $\T \indep \X\mid \X_{\calb}$. Definition of $\cala$ lead to the fact that $\calb \subseteq \cala$. Thus we only need to prove $\calb \supseteq \cala$ i.e. $(\T,\C) \indep \X\mid \X_{\calb}$.
%$\p(\T,\C\mid\X_{\calb})=\p(\C\mid\T,\X_{\calb})\p(\T\mid\X_{\calb})=\p(\C\mid\X_{\calb})\p(\T\mid\X_{\calb}),$
%where the second equation amounts to $\T \indep \C\mid \X_{\calb}$.
%$\p(\T,\C\mid\X)=\p(\C\mid\T,\X)\p(\T\mid\X)=\p(\C\mid\X)\p(\T\mid\X)=\p(\C\mid\X_{\calb})\p(\T\mid\X),$
%where the second and third equations result from $\T \indep \C\mid \X_{\calb}$ and $\C \indep \X\mid \X_{\calb}$ respectively. Because of $\p(\T\mid\X)=\p(\T\mid\X_{\calb})$ which results from $\T \indep \X\mid \X_{\calb}$, we have $\p(\T,\C\mid\X_{\calb})=\p(\T,\C\mid\X)$. Thus $\calb = \cala.$
%\eop

\noindent {\sc Proof of Proposition \ref{propgk}}.
Define $\H(\T^o,\widetilde \T^o\,\delta,\widetilde \delta)=E[2{\bSig}-(\X-\widetilde \X)(\X-\widetilde \X)\trans\mid \T^o,\widetilde \T^o, \delta,\widetilde \delta]$ and  let $H_{ii}(\T^o,\widetilde \T^o, \delta,\widetilde \delta)$ be its $i$th diagonal element. Following Proposition \ref{pro: Censored RD in CS}, we can obtain that
\begin{align*}
\spn[\H(\T^o,\widetilde \T^o, \delta,\widetilde \delta)]=\spn\{{\bSig}^{\frac{1}{2}}[2\I-\D(\T^o,\widetilde \T^o, \delta,\widetilde\delta)]{\bSig}^{\frac{1}{2}\trans}\}
=\spn({\bSig}\bfbeta_1,\ldots,{\bSig}\bfbeta_d).
\end{align*}
Thus $\H(\T^o,\widetilde \T^o\,\delta,\widetilde \delta)=\sum_{i=1}^d\lambda_i({\bSig}\bfbeta_i)({\bSig}\bfbeta_i)\trans$. Thus $H_{kk}(\T^o,\widetilde \T^o, \delta,\widetilde \delta)=0$ because the $k$th row of $({\bSig}\bfbeta_1,\ldots,$ ${\bSig}\bfbeta_d)$ is $(0,\ldots, 0)$.
Invoking Lemma \ref{Lemma: rewriten gkstar}, we can further derive that $g_{k}=E[H_{kk}(\T^o,\widetilde \T^o,\delta,\widetilde \delta)]^2/\sigma_k^4=0$. 
\eop

\noindent {\sc Proof of Proposition \ref{propgki}}.
Define $\H(\T^o,\widetilde \T^o\,\delta,\widetilde \delta)=E[2{\bSig}-(\X-\widetilde \X)(\X-\widetilde \X)\trans\mid \T^o,\widetilde \T^o, \delta,\widetilde \delta]$. Without loss of generality, set $e=1$ and $\calf=\{2,\ldots,k_0\}\supseteq \cala$. Define $p\times 1$ vector $\ba_{1\mid\calf}=(1,-{\bSig}_{\calf,1}\trans{\bSig}_\calf^{-1}, 0,\ldots,0)\trans$. Then we have $g_{1\mid\calf}=\sigma^{-4}_{1\mid\calf}E[\ba_{1\mid\calf}\trans \H(\T^o,\widetilde \T^o,\delta,\widetilde \delta)$ $\ba_{1\mid\calf}]^2$ by applying Theorem 2 in \citep{DR2007}. Similar with the proof of Proposition \ref{prop:gkstar}, $g_{1\mid\calf}=\sum_{ij}\sum_{lm}p_{ij}p_{lm}[\ba_{1\mid\calf}\trans \H_{ijlm}\ba_{1\mid\calf}]^2/\sigma^4_{1\mid\calf}$, where $\H_{ijlm}$ $=E[2{\bSig}-(\X-\widetilde \X)(\X-\widetilde \X)\trans\mid \delta=i,\T^o\in I_{ij},\widetilde\delta=l,\widetilde\T^o\in I_{lm})]$.
When $e=1 \in \cala$, condition (C4) guarantee there exists $k\in \{1,\ldots,d\}$ that $\beta_{ke}\neq 0$. Then $\ba_{1\mid\calf}\trans{\bSig}\bfbeta_k=
\sigma^2_{1\mid\calf}\beta_{ke}\neq0$. Since $\spn[\H_{ijlm}]=\spn({\bSig}\bfbeta_1,\ldots,{\bSig}\bfbeta_d)$, we have
\begin{align*}
\spn[\ba_{1\mid\calf}\trans \H_{ijlm}\ba_{1\mid\calf}]=\spn(\ba_{1\mid\calf}\trans{\bSig}\bfbeta_1,\ldots,\ba_{1\mid\calf}\trans{\bSig}\bfbeta_d).
\end{align*}
Thus there exists $a>0$ that $\ba_{1\mid\calf}\trans \H_{ijlm}\ba_{1\mid\calf}>a(\ba_{1\mid\calf}\trans{\bSig}{\bfbeta}_k)^2=a\sigma^4_{1\mid\calf}\beta_{ke}^2$ for at least one set of $(i,j)$ and $(l,m)$. Denote $\min p_{ij}$ by $p_{\text{min}}$. By condition (C5), we have $g_{e\mid\calf}=g_{1\mid\calf}>p_{\text{min}}^2[\ba_{1\mid\calf}\trans \H_{ijlm}\ba_{1\mid\calf}]^2$ $/\sigma^4_{1\mid\calf}>p_{\text{min}}^2a^2\sigma^4_{1\mid\calf}\beta_{ke}^4>c_0n^{-\kappa}$, where $c_0=p_{\text{min}}^2a^2c_2^2c_1^4$ and $\kappa=4\theta\le1/2$.
It follows that $g_{e\mid\calf}>2c_0n^{-\kappa}$ while $e\in \cala$.
\eop

\end{document}